\documentclass[preprint2]{aastex}

\newcommand{\lxlbol}{$L_{\rm X}-L_{\rm BOL}$}
\newcommand{\loglxlbol}{$\log(L_{\rm X}/L_{\rm BOL})$}

\shorttitle{Summary}
\shortauthors{Carina Survey}

\begin{document}

\title{GLOBAL X-RAY PROPERTIES OF THE O AND B STARS IN CARINA}

\author{Y. Naz\'e\altaffilmark{1,2}, P.S. Broos\altaffilmark{3}, L. Oskinova\altaffilmark{4}, L.K. Townsley\altaffilmark{3}, \and D. Cohen\altaffilmark{5}, M.F. Corcoran\altaffilmark{6}, N.R. Evans\altaffilmark{7}, M. Gagn\'e\altaffilmark{8}, A.F.J. Moffat\altaffilmark{9}, J.M. Pittard\altaffilmark{10}, G. Rauw\altaffilmark{1}, A. ud-Doula\altaffilmark{11}, N.R. Walborn\altaffilmark{12}}

\altaffiltext{1}{GAPHE, D\'epartement AGO, Universit\'e de Li\`ege, All\'ee du 6 Ao\^ut 17, Bat. B5C, B4000-Li\`ege, Belgium ; email: naze@astro.ulg.ac.be}
\altaffiltext{2}{Research Associate FRS-FNRS}
\altaffiltext{3}{Department of Astronomy \& Astrophysics, 525 Davey Laboratory, Pennsylvania State University, University Park, PA 16802, USA}
\altaffiltext{4}{Institute for Physics and Astronomy, Universit\"at Potsdam, 14476 Potsdam, Germany}
\altaffiltext{5}{Swarthmore College Department of Physics and Astronomy, 500 College Avenue, Swarthmore, PA 19081, USA}
\altaffiltext{6}{CRESST and X-ray Astrophysics Laboratory, NASA/GSFC, Greenbelt, MD 20771, USA ; Universities Space Research Association, 10211 Wincopin Circle, Suite 500 Columbia, MD 21044, USA}
\altaffiltext{7}{Harvard-Smithsonian Center for Astrophysics, 60 Garden St., Cambridge, MA 02138, USA}
\altaffiltext{8}{Department of Geology and Astronomy, West Chester University, West Chester, PA 19383, USA}
\altaffiltext{9}{D\'epartement de Physique, Universit\'e de Montr\'eal, Succursale Centre-Ville, Montr\'eal, QC, H3C 3J7, Canada ; Centre de recherche en astrophysique du Qu\'ebec, Canada}
\altaffiltext{10}{School of Physics and Astronomy, The University of Leeds, Leeds LS2 9JT, UK}
\altaffiltext{11}{Penn State Worthington Scranton, 120 Ridge View Drive, Dunmore, PA 18512, USA}
\altaffiltext{12}{Space Telescope Science Institute, 3700 Martin Drive, Baltimore, MD 21218, USA}

\begin{abstract}
The key empirical property of the X-ray emission from O stars is a strong correlation between the bolometric and X-ray luminosities. In the framework of the {\em Chandra} Carina Complex Project, 129 O and B stars have been detected as X-ray sources; 78 of those, all with spectral type earlier than B3, have enough counts for at least a rough X-ray spectral characterization. This leads to an estimate of the \lxlbol\ ratio for an exceptional number of 60 O stars belonging to the same region and triples the number of Carina massive stars studied spectroscopically in X-rays. The derived \loglxlbol\ is $-$7.26 for single objects, with a dispersion of only 0.21\,dex. Using the properties of hot massive stars listed in the literature, we compare the X-ray luminosities of different types of objects. In the case of O stars, the \lxlbol\ ratios are similar for bright and faint objects, as well as for stars of different luminosity classes or spectral types. Binaries appear only slightly harder and slightly more luminous in X-rays than single objects; the differences are not formally significant (at the 1\% level), except for the \lxlbol\ ratio in the medium (1.0--2.5\,keV) energy band. Weak-wind objects have similar X-ray luminosities but they display slightly softer spectra compared to ``normal'' O stars with the same bolometric luminosity. Discarding three overluminous objects, we find a very shallow trend of harder emission in brighter objects. The properties of the few B stars bright enough to yield some spectral information appear to be different overall (constant X-ray luminosities, harder spectra), hinting that another mechanism for producing X-rays, besides wind shocks, might be at work. However, it must be stressed that the earliest and X-ray brightest amongst these few detected objects are similar to the latest O stars, suggesting a possibly smooth transition between the two processes. 

\end{abstract}


\section{INTRODUCTION}
The existence of X-ray emission from hot, massive stars was predicted three decades ago by \citet{cas79} and it was serendipitously discovered at the same time during early observations by the {\em Einstein} satellite; massive O stars in the Carina Nebula were among the first detections \citep{sew79}. It was immediately found that the X-ray emission of these O stars was proportional to their optical luminosity \citep{har79}.  The relation was then refined to be proportional to their bolometric luminosity (\citealt{pal81}; as the first observations dealt with stars of similar types and reddening, both relations were equivalent).  X-rays are now thought to be generated in wind shocks and, as the winds are line-driven, it may seem quite natural that a relationship between X-ray and bolometric luminosities exists. However, the theoretical derivation of this relationship is not obvious: \citet{owo99} showed that, while the X-ray luminosity ``naturally'' scales with the wind density parameter $\dot M / v_{\infty}$, it only scales with the bolometric luminosity if there is ``a delicate balance between X-ray emission and absorption'' and ``a special form for the radial distribution of wind shocks''. A better knowledge of this relationship may thus yield a better understanding of the X-ray emission of O stars. Finally, {\em Einstein} observations also showed that the detection rate of B stars was much lower than that of O stars \citep{gri92}. 

The X-ray properties of O and B stars were constrained more accurately by \citet{ber97} using the {\em ROSAT} All-Sky Survey (RASS). Using 237 detections, Bergh\"ofer et al.\ confirmed the decline in the detection rate towards later spectral types (all stars with type O7 or earlier were detected as X-ray sources, while at most 10\% of B3--B9 stars were detected). This fact, correlated with a higher incidence of variability and binarity among the latest types, led to the conclusion that low-mass pre-main sequence (PMS) companions could be responsible for the X-ray emission of late-B stars. Bergh\"ofer et al.\ also confirmed that the \loglxlbol$\sim$--7 relation applies down to $\log(L_{\rm BOL})\sim38$~erg~s$^{-1}$ (corresponding approximately to spectral type B1III--V) although with a large dispersion ($\sigma$ of 0.4\,dex, or a factor of 2.5). \citet{coh97} further showed a steep decrease in X-ray luminosity (with \loglxlbol\ reaching --8.5 at B2) and a softening of the emission for fainter stars. At the same time, studies of O+OB binaries revealed enhanced X-ray luminosities for these objects, attributed to the collision between the two stellar winds \citep{chl89}. No such enhancement was reported for B+B binaries (even early-B binaries) but such systems were much less studied than O+OB binaries.

In recent years, many star clusters were observed by {\em XMM-Newton} or {\em Chandra}, often for detailed studies of the PMS population or the diffuse emission \citep[e.g.][]{gue07,tow03}. Only a few of these clusters harbor a significantly large number of O stars, which is required for statistical studies of their X-ray properties. The O star data also often had limited signal-to-noise, preventing a full spectral analysis of the X-ray emission. In the favorable cases, estimates of the X-ray luminosities appeared compatible with the \loglxlbol$\sim$--7 relation, although a large scatter was often present due to limited knowledge of the stellar content \citep[][]{naz08,wan08}. Some rare, peculiar objects were overluminous in X-rays, as exemplified by the Orion Trapezium \citep{ste05}. Only in a few cases could in-depth studies of the \lxlbol\ relation be performed. These analyses relied on a precise knowledge of the cluster's stellar content and detailed stellar properties (multiplicity, reddening, bolometric luminosity, and X-ray spectrum). They revealed a rather tight \lxlbol\ relation (e.g.\ NGC~6231, \citealt{san06}, and Car~OB1, \citealt{ant08}). Only a few O+OB binaries were found to be truly overluminous \citep[see also][]{osk05}; X-ray-bright wind-wind collisions thus now appear to be the exception rather than the rule, although a detailed physical explanation is still lacking. 

However, the RASS and cluster samples differed in many ways \citep[see][ for a full discussion]{naz10}, notably regarding the number of targets analyzed ($>$200 vs.\ $\sim$20), their homogeneity (a mix of different clusters and field stars for the former vs.\ a single cluster/association for the latter) and the data quality (count rates and hardness ratios vs.\ detailed spectral analyses). The first attempt to combine both approaches was recently performed by \citet{naz09}. This global spectral analysis of the massive stars detected in the 2XMM catalog confirmed the lack of overluminosities in binaries as well as the large scatter around the \lxlbol\ relation for an inhomogeneous population, hinting at differences between clusters.

This very large {\em Chandra} program targeting the Carina Complex now constitutes a new opportunity to look at the X-ray properties of a large population of massive stars, with better homogeneity between stars than in the 2XMM sample. This paper aims at more than tripling the number of massive stars analyzed in Carina in the X-ray domain compared to what is available at the present time \citep{ant08}. Section 2 will summarize the data characteristics, Section 3 presents the derived \lxlbol\ relation, and Sections 4, 5, 6, and 7 discuss the hardness of the spectra, the properties of binaries, the weak wind stars, and the faintest objects, respectively. Finally, Section 8 summarizes our results.

\section{THE DATA}

The {\em Chandra} Carina Complex Project (CCCP) is described in detail in \citet{Townsley11}. The source detection process and spectral extractions are described by \citet{Broos10} and \citet{Broos11}.  Broos et al.\ also describe the complicated completeness limits of the survey, which vary across the field due to {\em Chandra}'s changing point spread function, vignetting, and the spatially-complex exposure times across the mosaic.

Parallel to the X-ray data analysis, a list of 70 O-type and 130 B-type stars in the CCCP field of view was compiled from \citet[see also \citealt{Gagne11}]{ski09}, a catalog of stars that have been studied spectroscopically in the literature.  We concentrate on these OB stars in this paper because their spectral types are known.  A discussion of candidate O and B stars, derived from photometric data, is given in \citet{Povich11} and \citet{Evans11}. Note that all but 3 of the 130 B-type stars have spectral types earlier than B3; they can thus be considered as hot, massive objects.  The 3 later-type B stars are undetected in X-rays; these 3 sources are not discussed thoroughly in this paper.

For a distance of 2.3\,kpc, the bolometric luminosities of the catalog objects were generally derived from the known spectral types and SED fitting \citep{Povich11} using UBVJHKs magnitudes and Spitzer-IRAC photometry. This ensures that the derived values of the bolometric luminosities do not depend greatly on the choice of the extinction law; other systematic errors (due e.g. to the choice of atmosphere models) are $<$5\% \citep{Povich11}. In only 17 cases bolometric luminosities were derived from V-magnitudes and bolometric corrections due to the lack of reliable photometry \citep{Povich11,Gagne11}.  Comparisons between the results of the two methods revealed only limited differences ($<$0.2\,dex) between the bolometric luminosities. Povich et al.\ also derived reddenings from the photometry, leading to estimates of the interstellar absorbing column $N_H(ISM)$. To this aim, we use a gas-to-dust ratio $N_H/A_V=1.6 \times 10^{21}$\,cm$^2$\,mag$^{-1}$ \citep{vuo03,get05} and $R_V=4$. Such a value of $R_V$ is adapted to the choice of a 2.3\,kpc distance \citep{wal95}; it also provides the best fit for the photometry of O and B stars \citep{Povich11}. Photometric uncertainties and star-to-star variations in extinction yield errors $<$20\% in bolometric luminosities \citep{Povich11}, but we acknowledge that a large scatter exists around the total-to-selective absorption, the used value being only the best currently available. The binary status of 15 O-type stars was obtained from \citet[spectroscopic binaries]{rau09} and \citet[mostly visual binaries]{nel04}. Note that we used the same spectral type for QZ~Car as \citet{Parkin11}. Four additional B stars are catalogued as binaries, because they appear as such in \citet{nel04} or are listed as B+B in the \citet{ski09} catalog used for collecting spectral types.

This massive star catalog was first cross-correlated with the CCCP list of X-ray sources \citep{Broos11} and then used to directly explore the X-ray data at the positions of the hot stars \citep{Gagne11}. 
In total, 129 matches were found (Table~\ref{repart}), or about 65\% of the objects in the Skiff-based catalog of OB stars (see above).
Carina's three Wolf-Rayet stars and the luminous blue variable $\eta$~Carinae were also detected in the CCCP; they are mentioned in \citet{Townsley11}. Two O stars, 66 early-B stars, and the 3 late-B stars mentioned above remain undetected in the CCCP. 
This is consistent with the average completeness limit of this survey, $\log L_{\rm X}\sim30$ \citep{Broos11}. Indeed, with a typical \loglxlbol\ of --7 and minimum bolometric luminosities of $\log L_{\rm BOL}\sim38.2$, all O-type stars are expected to be detected with such a limit and even to display enough counts for at least a rough spectral analysis. On the contrary, most B stars have much lower bolometric luminosities and \loglxlbol\ ratios.  The low detection rate in their case is therefore unsurprising; only the earliest-type and brightest objects could thus be analyzed here (see \citealt{Evans11} for more details on the faintest, undetected B stars). 

\subsection{Spectral Fitting}

X-ray spectra were automaticaly extracted for all X-ray sources. Whenever several observations of the same target were available, the individual spectra were merged (but note that the far off-axis observations of QZ~Car reported by \citealt{Parkin11} were not included in this study). Spectral fitting was done on the grouped spectra (grouped such that each spectral bin displays at least a signal-to-noise ratio of 3) of the 78 O and B sources which displayed at least 50 net counts. We arbitrarily split the sources in two groups: reliable if the source has at least 100 counts (which is the case for 29 single O stars, 13 O+OB binaries, and 10 B stars, see Table~\ref{repart}), and somewhat reliable if the source has 50--100 counts (true for 16 single O stars, 2 O+OB binaries, and 8 B stars). Errors are indeed larger for objects of the second group (see Table~\ref{bestfit}). The fitted models were of the form $tbabs_1 \times tbabs_2\times \sum_1^{1or2}apec$. The $tbabs$ components represent neutral absorptions \citep{wil00}; the first one is fixed to the intervening interstellar column and the second one accounts for potential additional, circumstellar absorption.  The small number of counts and the limited sensitivity at low energies prevented us from using more sophisticated wind absorption models. The last part of the model corresponds to the emission from one or two optically-thin thermal plasmas \citep{smi01}. The second thermal plasma component was added only if necessary, i.e., if it led to a significant improvement in the $\chi^2$ of the fit. The abundances were always kept to solar because the sources generally have too few counts to ascertain abundance variations. 

Table~\ref{bestfit} provides the fitting results. Column (1) shows the X-ray source name, while Columns 2--4 give the observed basic source properties:  the number of net counts in the 0.5--8.0\,keV band together with their associated 68\% confidence intervals, the photon flux (net counts divided by the mean effective area and the exposure time), and the median energy of the recorded counts \citep{Broos11}. The next columns provide the fitting results themselves: the reduced $\chi^2$, the number of degrees of freedom, the absorbing columns (interstellar then circumstellar), the temperatures and emission measures of each thermal component, and finally the observed fluxes in the 0.5--10.0\,keV energy band (these energy boundaries were chosen to ease comparison with other, previous \lxlbol\ studies).   For each fitted parameter, the lower and upper limits of the 90\% confidence intervals are listed as indices and exponents, respectively (if the limit is not explicitely noted, the parameter should be considered as unconstrained in that direction). Note that for the fluxes, these errors (estimated using the Xspec command {\it flux err}) are only indicative as they do not fully or correctly take into account the spectral model uncertainties and/or the correlations between spectral parameters.   

\begin{figure}
\includegraphics[width=9cm, bb=25 270 590 710, clip]{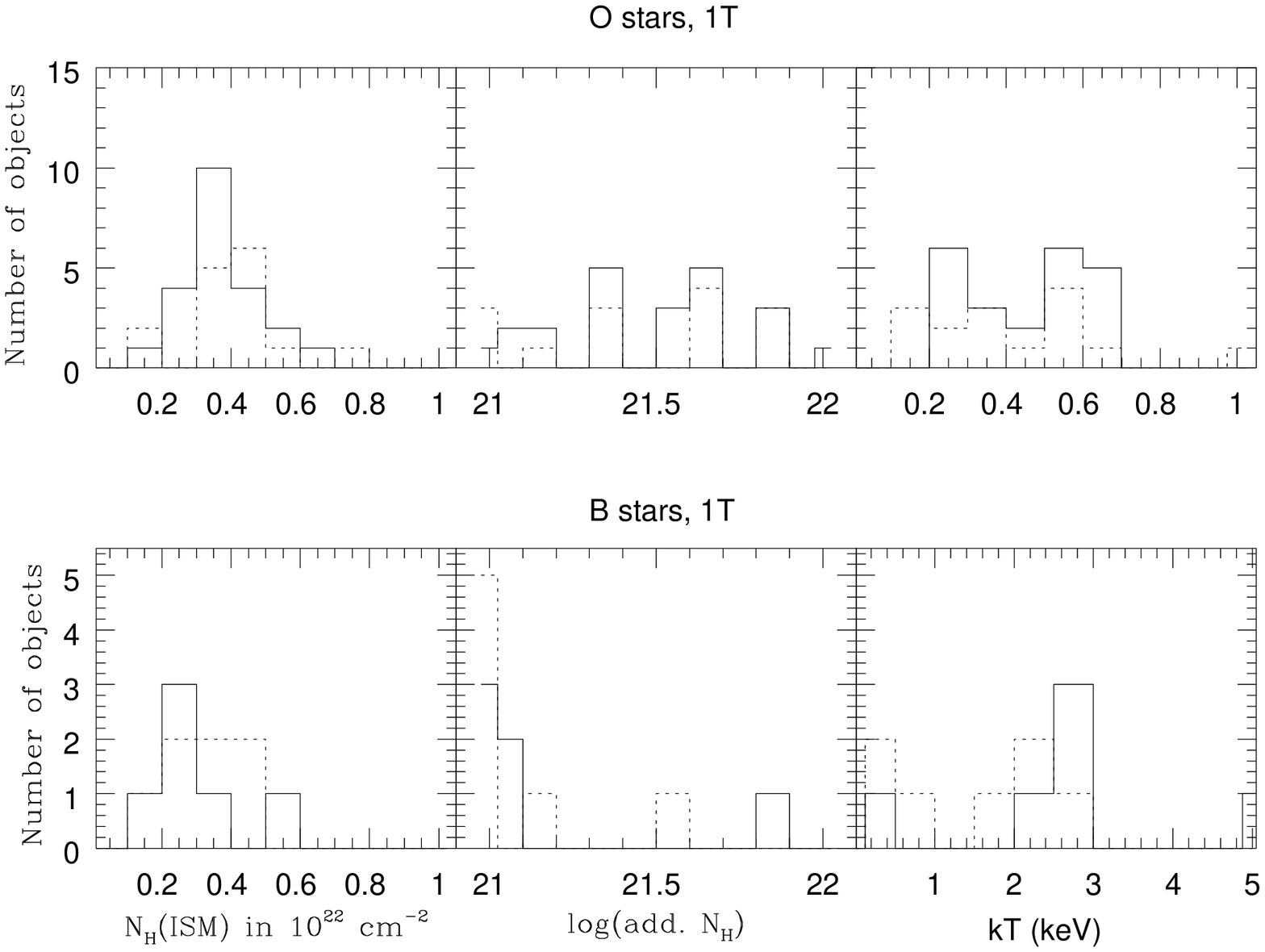}
\includegraphics[width=9cm]{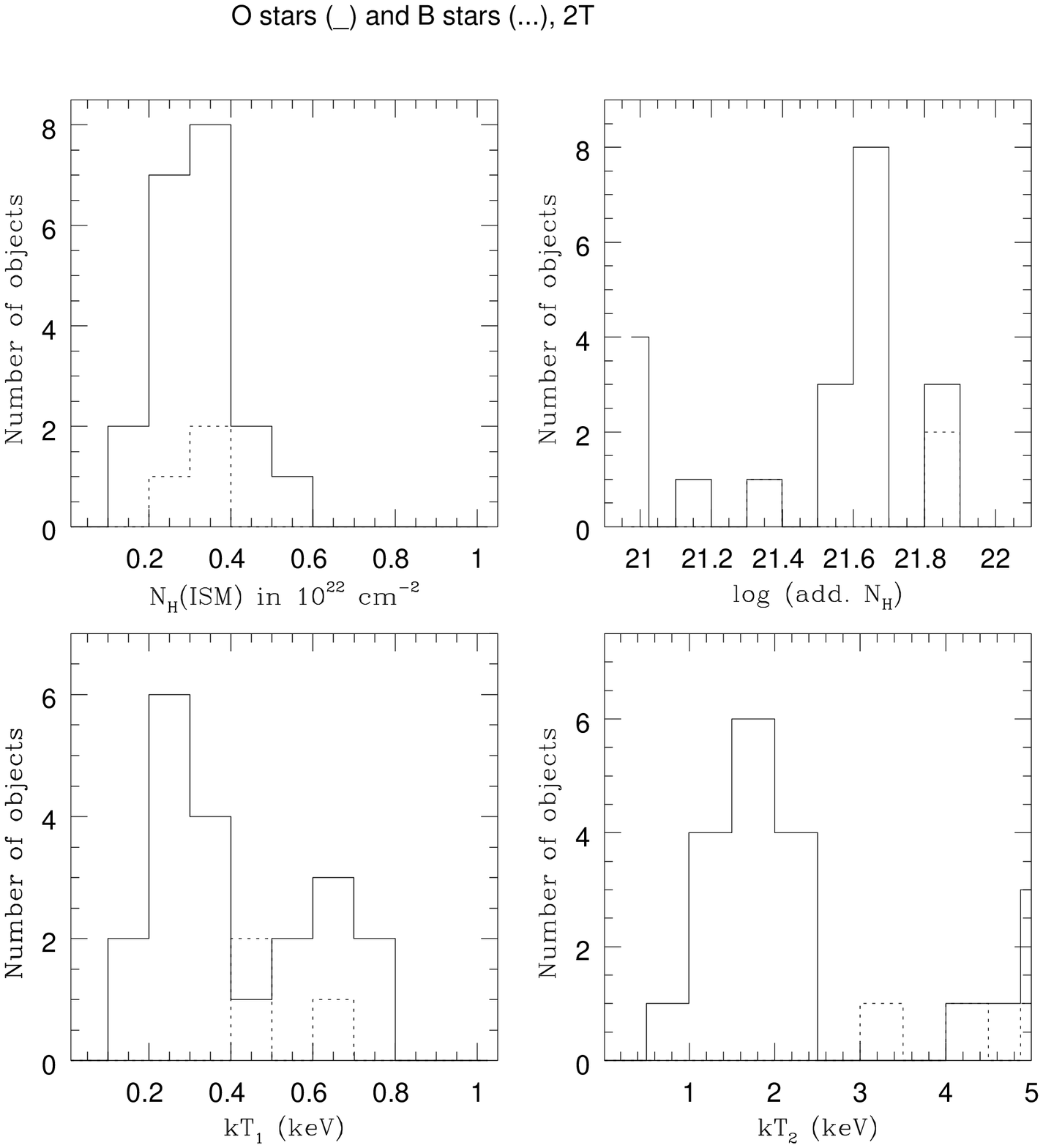}
\includegraphics[width=5cm]{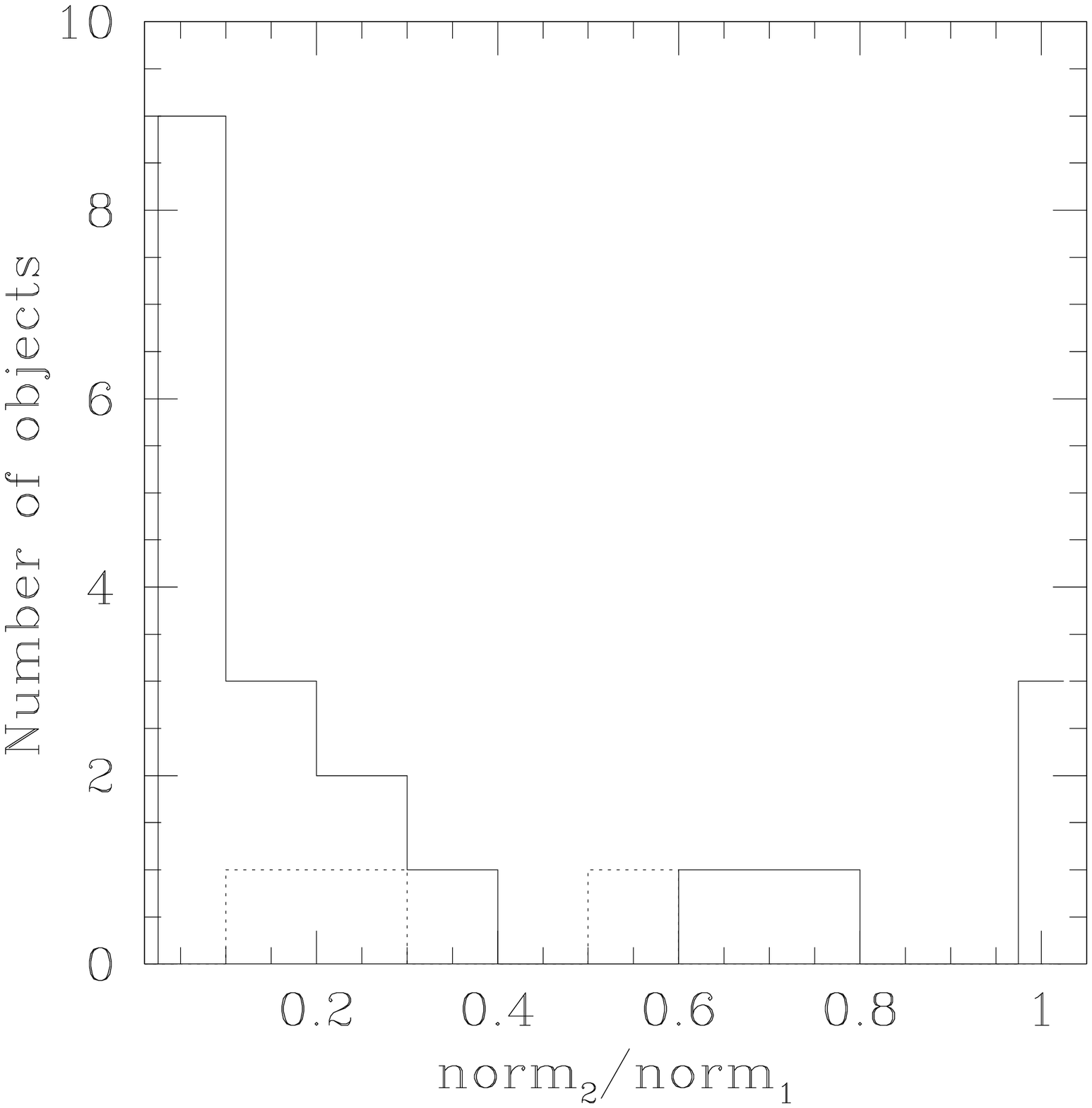}
\caption{Distribution of the spectral parameters for O and B stars. For the top six subfigures, the solid line refers to sources with $>$100 counts and the dotted line to sources with 50--100 counts.  For the bottom five subfigures, only the sources with $>$100 counts were considered, with O stars shown using a solid line and B stars as a dotted line. \label{fig:hist}}
\end{figure}

Histograms of the temperatures and absorbing columns were derived from the best-fit values (Figure~\ref{fig:hist}). The results are similar to those of \citet{naz09}: for O-type stars, the favored $kT$ is about 0.2--0.6\,keV, a second temperature of about 2\,keV is sometimes necessary, and there appears to be a significant additional (circumstellar) absorbing column of about 4$\times$10$^{21}$\,cm$^{-2}$.  When two temperatures are needed to fit a spectrum, the hotter component is of reduced strength compared to the main, ``cool'' component. Note that the value of the additional absorbing column does not show any dependence on the bolometric luminosity of the associated star. For B-type stars, no additional absorbing column is needed, but the dominant temperatures are higher.

Due to their high count rates, photon pile-up may affect four sources in the O and early-B population \citep{Broos11}: HD\,93129A, HD\,93205, HD\,93206, and HD\,93250. We ignore the mild pile-up of HD\,93403. In this paper, the spectral fitting for these sources was performed on reconstructed spectra (see Broos et al.\ for details on the reconstruction method). Additional fittings performed on individual X-ray spectra of these bright objects are presented in \citet[QZ~Car]{Parkin11} and \citet[other massive stars]{Gagne11}.

\subsection{Comparison with Previous Studies}
\begin{figure}
\includegraphics[width=9cm]{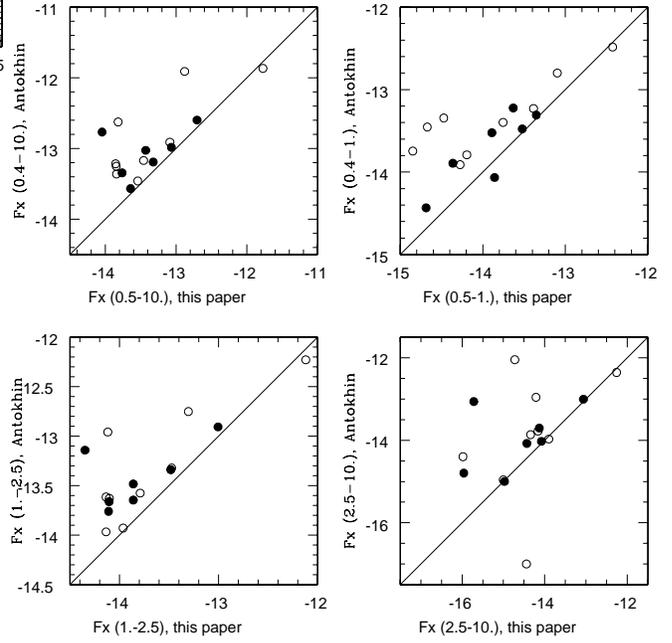}
\caption{Comparison of the observed (absorbed) fluxes between \citet{ant08} and this paper. Filled circles correspond to X-ray sources with companions within 5$\arcsec$ and open circles to isolated X-ray sources. \label{fig:igor}}
\end{figure}

Ten sources are in common with the {\em Chandra} analysis reported by \citet{eva04}; except for the two known variable stars (HD\,93250, see \citealt{rau09}, and Tr16\,MJ\,496, described below), their observed fluxes agree with ours to within the errors (which are of the order 10--20\%). For completeness, we can also compare our results with the {\em XMM-Newton} study of \citet{ant08}. Sixteen of their stars (excluding the O+OB binaries and close pairs such as HD\,93129AB and HD\,93161AB) are in common with our survey. However, there is not an exact one-to-one scaling when comparing the observed fluxes of these sources (see Figure~\ref{fig:igor}). This difference appears especially important for 3 objects:  HD\,303308 (O4V), Tr14\,MJ\,127 (O9V), and Tr14\,MJ\,181 (B1.5V). These flux differences may be explained in several ways: energy band boundaries, crowding, variability, and exposure lengths. First, \citet{ant08} uses a soft boundary of 0.4\,keV while we used 0.5\,keV. This can explain why Antokhin's fluxes are systematically higher than ours (but this cannot explain a factor of 2 difference). Also, {\em Chandra}, which has better spatial resolution than {\em XMM-Newton}, can distinguish close neighbors more easily, again making the {\em XMM-Newton} X-ray luminosities systematically overestimated. For example, HD\,93129AB was seen as a single source in {\em XMM-Newton}, while the two components are here clearly separated, with additional, fainter close companions detected. However, crowding is not the only cause for the difference, as sources without any detectable companions are still much more luminous in Antokhin's data (e.g.\ HD\,303308, Tr14\,MJ\,181). Variability may also cause brightness differences. If the X-ray emission of the B-star Tr14\,MJ\,181 is magnetic in origin (X-rays from a PMS companion or a corona intrinsic to the star), flares should be quite common, though none is detected in our data.  Finally, the exposure and sensitivities are quite different in the {\em XMM-Newton} and {\em Chandra} datasets. On average, the exposure and sensitivity were lower for {\em Chandra}; the low number of counts certainly explains the scatter of the hard flux (in the 2.5-10.0\,keV band) but this can also be true for a few particularly faint objects (e.g.\ Tr14\,MJ\,127, LS\,1897). There is thus no single explanation for the observed flux differences between Antokhin's work and ours. 

Comparing the final X-ray luminosities (corrected for interstellar absorption) and the associated \lxlbol\ ratios to the values from \citet{ant08} is even more complicated. First, the interstellar columns are different, as we did not choose the same gas-to-dust ratio and $R_V$; with a larger assumed distance (increasing luminosities by 20\%) and interstellar columns increasing by 15\% in Antokhin et al.\ because of the changed gas-to-dust ratio alone, the absolute values of the absorption-corrected X-ray luminosities from Antokhin et al.\ are expected to differ from our values. Second, the bolometric luminosities were derived in a very different way, also leading to differences. As a result, Antokhin et al.\ find an average \loglxlbol=--6.58, i.e., a value larger by about 0.7\,dex compared to that derived from our data (see next section). 

It is interesting to note that, using $R_V=3.1$ and deriving the bolometric luminosities from V magnitudes, our data yield \loglxlbol=--6.99. The choice of a given calibration therefore clearly affects the absolute value of the \lxlbol\ ratio and one should not simply compare the values of these ratios given in different studies. Only homogeneous studies of global samples, such as done in \citet{ber97} or \citet{naz09}, can be used for such purposes. However, as our dataset is homogeneous in itself (same instrument and similar completeness limit throughout the field-of-view), its results can be seen as self-consistent and a comparison between different objects studied in this paper is feasible. 

\section{The \lxlbol\ Relation} 

Table \ref{lxlbollist} summarizes the main properties of the brightest sources (those with $>$50 recorded counts). The first two columns give the X-ray source name and the stellar identification. They are followed by the spectral type (Column 3), the binary status (Column 4), the bolometric luminosity (Column 5 - columns 2--5 are reproduced from the stellar catalog by \citealt{Povich11, Gagne11}), the X-ray luminosities in four energy bands (defined below, Columns 6--9), and the \lxlbol\ ratios in the same energy bands (Columns 10-13). Note that the X-ray luminosities were corrected by the interstellar absorption only (this enables us to compare the actual X-ray throughput of massive stars and eases the comparison with other studies, as the circumstellar absorption is not always taken into account in the same way).  The luminosities are given in the 0.5--10.0\,keV (``total''), 0.5--1.0\,keV (``soft''), 1.0--2.5\,keV (``medium''), and 2.5--10.0\,keV (``hard'') energy bands. Most of the X-ray emission of massive stars is emitted in the soft and medium bands, the former being more sensitive to absorption effects; the relative strength of these two bands is also an indication of the softness of the emission. Hot stars generally emit few X-rays in the hard band and the corresponding fluxes therefore show large uncertainties. However, this band is useful for detecting peculiar phenomena such as magnetically-confined winds or colliding winds (in which cases the hard X-ray flux is enhanced). These phenomena will be explored below.

Averages ($X=\sum X_i/N$) and dispersions ($\sigma=\sum (X_i-X)/(N-1)$) of the \loglxlbol\ ratio were evaluated for several different samples of objects. No weighting was applied and comparisons of the means were made following the methods outlined in \citet[Chapter 7.3]{lin68} and for a significance level of 1\%. Table~\ref{ratio} and Figure~\ref{fig:lx} summarize the results. We emphasize again that the X-ray luminosities quoted here are corrected only for the interstellar absorbing column and that the hard X-ray fluxes are unreliable due to the small count rates at such high energies for most of our X-ray sources.

\subsection{O Stars}
As usual, the O-type stars show a clear \lxlbol\ correlation, with similar results regardless of the sample chosen (i.e., $>$100 counts vs.\ 50--100 counts). Only 3 stars strongly deviate from the average behavior (Figure~\ref{fig:lx}): HD\,93250, HD\,93403, and Tr16\,MJ\,496 (also known as Tr16\,22). The first star is a suspected binary, due to the presence of non-thermal radio emission theoretically expected to arise in wind-wind collisions and to the detection of large X-ray variations \citep{rau09}. However, the signature of the companion has never been directly detected \citep[see][and references therein]{rau09}. The second star is a known binary, whose X-ray emission, monitored by {\em XMM-Newton}, shows clear signs of a wind-wind collision \citep[see][for a full analysis of the phenomenon]{rau02}. 

The third object was also identified as overluminous (and variable) by \citet{ant08}, and its emission appears quite hard for an O star (see below), suggesting a mechanism other than the usual wind-shock process. In such cases, two possibilities exist: wind-wind collisions and magnetic wind confinement. Although radial velocity variations may have been detected \citep{combi}, the former mechanism is not favored since such a large overluminosity ($>$1\,dex) is neither observed nor theoretically expected in late-type O+OB colliding wind binaries. Tr16\,MJ\,496 could thus be a magnetic object that might resemble $\theta^1$\,Ori\,C \citep{gag05}; additional spectropolarimetric data and multiwavelength monitoring are needed to confirm this hypothesis. In summary, in all three cases, the detected overluminosities at high energies confirm previous observations and can be explained by X-rays generated in colliding winds or magnetically-confined winds in addition to the intrinsic, wind-shock emission. 

Discarding these 3 sources from the \loglxlbol\ average yields a final \loglxlbol\ of $-$7.26 and a dispersion of only 0.2\,dex (Table~\ref{ratio}), quite similar to that found in NGC~6231 once its two binaries with X-ray-bright wind-wind collisions have been discarded \citep{san06}. This confirms that the \lxlbol\ relationship is quite tight for a homogeneous population of stars; the observed scatter in \loglxlbol\ found from large samples \citep{ber97,naz09} is thus clearly due to the inhomogeneity of the stellar populations in these samples. To study further the \lxlbol\ relationship, we have compared different stellar groups (Table~\ref{ratio}). There are no clear, significant differences (at the 1\% level) in \lxlbol\ ratios for stars of different luminosity class, as was notably claimed by \citet{alb07} for Cyg~OB2 (considering however that the number of O supergiants is too low for providing statistically meaningful results).  Neither were there significant differences (at the 1\% level) in \lxlbol\ ratios for stars of different spectral types or of different luminosities (except for the faint vs.\ medium-bright objects in the medium energy band). Compiling single O stars belonging to the clusters Tr14 and Tr16/Cr232 using \citet{cud93}, Table~1 from \citet{wal95}, Table~3 from \citet{deg01}, and Tables~5 and 7 from \citet{car04}, a comparison of \loglxlbol\ between these two stellar groups can be attempted. Single stars of Tr16 yield very similar results as single stars of the whole Carina region, whereas single stars of Tr14 display a systematicaly higher \loglxlbol\ (as well as smaller dispersions in the total and soft bands) than stars of Tr16 or of the whole area (Table~\ref{ratio}). However, the difference in \loglxlbol\ is not formally significant (at the 1\% level) and only hints at potential differences between clusters, as was already reported in \citet{naz09}.

\subsection{B Stars}
On the other hand, the B stars detected with more than 50 counts show no strong correlation between X-ray and bolometric luminosities (Figure~\ref{fig:lx}) and larger dispersions than for O stars are measured in all bands (Table~\ref{ratio}). In all cases, the \lxlbol\ ratios are larger than that of O stars, especially for the medium energy band (e.g. \loglxlbol = --6.9..--7.4 vs --7.84 in that band). A possible trend towards higher soft X-ray luminosities for more luminous objects is seen in the soft band (with a similar slope as for O-type stars). In contrast, the medium-band X-ray luminosities appear rather constant, with even a shallow decrease towards higher bolometric luminosities (Figure~\ref{fig:lx}). A similar situation is seen at the highest energies, although there are now a few outliers (most probably due to the low number of hard photon counts recorded for these objects). However, only 17 B stars were bright enough to be used for the \lxlbol\ analysis, while 45 single O stars spanning the whole range of bolometric luminosities were available in the previous section. The poor statistics definitely require a confirmation of our results with a larger number of objects (see e.g.\ \citealt{naz09}), notably those of lower luminosity, where significant incompleteness biases can arise.

\section{ADDITIONAL CORRELATIONS} 

We also checked for additional correlations besides how \lxlbol\ varies with different stellar subgroups. We remind the reader once again that the X-ray luminosities quoted here are corrected only for the interstellar absorbing column, that the hard X-ray fluxes are unreliable due to the small count rates at such high energies for most of our X-ray sources, and that only few B stars had enough counts to be analyzed spectroscopically.

\subsection{\loglxlbol\ vs.\ $L_{\rm BOL}$}
First, we investigated the \loglxlbol\ dependence on the bolometric luminosity (Figure~\ref{fig:ratio}). For O stars, there is essentially no trend; the \loglxlbol\ relation is thus well-constrained to a constant, as seen in the previous section. 

One can also look at the evolution of the dispersion around that constant, in the case of single objects but excluding the well-known problematic cases of HD\,93250 and Tr16\,MJ\,496. The dispersions for early-type (O2--O5.5), mid-type (O6--O8), and late-type (O8.5--O9.7) stars were compared using F-tests. While early- and mid-type stars display very similar dispersions, that of the late-type stars appears significantly different (at the 1\% level), but only in the medium energy band. It could be interpreted as a hint that the weaker stellar winds of late-type stars could be more affected by magnetic confinement than those of early- and mid-type stars\footnote{Using the stellar properties of \citet{mar05b} for O5.5, O7, and O9.5 main-sequence stars, a mass-loss rate derived from these parameters using the recipe of \citet{vin00}, and a terminal velocity of 2000\,km\,s$^{-1}$, the magnetic confinement parameter $\eta$ \citep{udd02} reaches the critical unity value for magnetic fields that are 35\,G for late-O stars and four times larger, 150\,G, for early-type stars: it is thus easier to confine the wind from late-type stars. Similar conclusions are reached when comparing O9.5 supergiants, giants, and main-sequence objects; the magnetic field requirements in this case are 35\,G for the latter and 60\,G for the former. Note that we did not attempt a refined test, splitting stars into groups of similar spectral types {\em and} luminosity classes since the O stars in Carina are too few in number to provide a meaningful result in that case.}; harder, additional X-rays could then contribute to the total high-energy emission when the magnetic field is sufficiently strong, leading to a larger scatter in the medium energy band. Since the stellar winds of hot stars are well-known to be radiation-driven, looking at the dispersions as a function of bolometric luminosity should confirm the above trend (i.e., lower luminosity objects should display a larger dispersion in their \loglxlbol). This is not the case, however, as the difference totally disappears:  using two luminosity bins, quasi-identical results are found for stars with $\log(L_{\rm BOL})$ above and below 38.5; using three luminosity bins, the largest dispersion is only detected for the total band and the brightest stars ($\log[L_{\rm BOL}]>38.75$). With such contradictory results, we cannot claim to detect a larger impact of the magnetic confinement on the late-O stars of lower luminosity. This correlates well with the findings by polarimetric surveys of only a few O stars with strong magnetic fields, hence few objects with winds affected by magnetic confinements (see e.g. \citealt{gue09} and references therein).

For the B stars detected with more than 50 counts, a tight relationship is seen between \loglxlbol\ and the bolometric luminosity, especially in the total and medium energy bands (Figure~\ref{fig:ratio}). Correlation coefficients in those energy bands reach at least $-$0.91, and the linear fits yield slopes compatible with minus unity. This may simply reflect the fact that the X-ray luminosity of these B stars appears rather constant with respect to the bolometric luminosity, as already noted in the previous section. The average (log) X-ray luminosities (and their associated dispersions) are 31.16$\pm$0.13 in the total band and 30.68$\pm$0.20 in the medium band considering all single stars and 31.21$\pm$0.14 and 30.78$\pm$0.17, respectively, considering only the single B stars with $>$100 cts. We caution however that such luminosities are close to the brightness limit for spectral analysis (the average detection limit is $\sim10^{30}$\,erg\,s$^{-1}$, and 50 counts are needed for a rough spectral analysis). Clearly, a more extensive and sensitive study is needed to confirm the nearly constant luminosity of these B stars (see also Section~7 below).

As mentioned in the introduction, the X-ray emission of B stars, since it is observed only for a small fraction of objects, may not be linked to the B-star itself but to a close PMS object, either a physical companion or a line-of-sight coincidence in dense clusters \citep[see also][ for more on this subject]{Evans11}. Indeed, when flaring, such PMS stars often reach 10$^{30}$\,erg\,s$^{-1}$ and even in some rare cases 10$^{31}$\,erg\,s$^{-1}$ \citep{gue09}. The observed luminosity of the X-ray-brightest B stars in Carina is thus compatible with the maximum luminosity of flaring PMS stars. Since such luminosities are only reached during flares, we should detect a strong variability of our sources if their X-ray emission is due to PMS flaring, but the lack of counts prevents us from deriving lightcurves for all objects.  Also we do not see any clear connection between a high X-ray luminosity and flares even when flares are detected (for more details, see \citealt{Gagne11} and \citealt{Evans11}). In addition, the hard spectra of B stars appear compatible with the PMS scenario but in some peculiar B stars, magnetic phenomena can also produce hard X-ray emission \citep[see the case of $\sigma$\,Ori\,E ][]{san04,udd06}. Finally, it should be noted that a gradual transition of behaviors is observed between late-O stars and the brightest, earliest B stars (see below). This could therefore advocate for an intrinsic source of X-rays in (at least some) early B stars, such as embedded-wind shocks for the earliest ones and magnetically-confined winds for some others.

\subsection{Hardness Ratios}
A second analysis involved the hardness ratio, defined as the ratio of the medium to soft (unabsorbed) X-ray luminosities (Figure~\ref{fig:hr}). In the B stars, this ratio appears at 0.3--0.4\,dex for most objects; it corresponds to the ratio of a plasma with a temperature of $>$1.5\,keV in the absence of absorption. This value is clearly above the average ratio for O stars, but it should be noted that the brightest objects have a lower ratio which is quasi-identical to that of late-O stars. Looking at this ratio as a function of spectral type or luminosity class confirms the trend. The earliest B stars (B0) display low hardness ratios (i.e., the X-ray emission is soft) compared to their slightly later (B1--B2) counterparts; a similar trend is seen when comparing supergiant B stars with their main-sequence counterparts. Note however that the number of B stars bright enough to yield a usable spectrum in the CCCP is low, and that this may blur the trends -- indeed, the three supergiant B stars are all of type B0.

Concerning O stars, there appears to be no trend with luminosity class, and the relation with spectral type appears scattered (the only conclusion being that low ratios are not favored in the case of the earliest O stars). However, an overall shallow trend appears when hardness ratio is plotted versus the bolometric luminosity. Indeed, except for a few faint stars (for which the 1.0--2.5\,keV fluxes are uncertain, see crosses in Figure~\ref{fig:hr}) and the peculiar Tr16\,MJ\,496, harder X-ray emission seems to correlate loosely with higher bolometric luminosities. However, the correlation coefficient is far from being significant due to the large scatter. In the framework of the wind-shock model, there may be two possible causes for this shallow trend. First, this could be an absorption effect. The brightest stars have stronger, hence more dense, stellar winds, which can lead to a larger intrinsic absorption and therefore a harder appearance of the spectrum. Our fits provide estimates of the circumstellar absorption, but its correlation with hardness ratio again seems quite shallow and scattered, thus not formally significant (Figure~\ref{fig:hr2}), and as already mentioned in Section~2.1, there is no trend whatsoever between this additional column and the bolometric luminosity. Second, \citet{wal09} detected (1) a larger ionization of the plasma in early-type stars (implying a hotter plasma) and (2) a shift of the overall spectral distribution towards lower energies for late-type stars. These trends, if confirmed, would also be compatible with a higher hardness ratio towards larger bolometric luminosities. Without detailed hydrodynamic simulations and better statistics, it is still unclear whether the hotter plasma, the higher absorption, or a combination of both effects is the cause of this shallow trend.

Finally, as the X-ray emission from hot stars originates from their stellar winds, correlations of the hardness ratios with wind parameters should be investigated. As reliable, homogeneous mass-loss rates are unavailable for our sample stars, we used only the terminal wind speeds from \citet{how97}. Ten O-type stars are in common between Howarth's catalog and this survey. Again, a shallow trend (in the sense of harder emission for faster winds, see Figure~\ref{fig:hr2}) may be present but the scatter is large and the trend detected by eye is clearly dominated by the two extreme stars, HD\,93027 (low $v_{\infty}$, low hardness ratio) and HD\,93129A (high $v_{\infty}$, high hardness ratio), casting doubt on its existence. 

In any case, what seems obvious from the different tested relationships is that the brightest and earliest B stars behave like the faintest and latest O stars and that there is a gradual transition towards B stars of later types (B1--B2). This continuum of behaviors could be explained by wind-shock emission essentially vanishing below bolometric luminosities of about 10$^{38}$\,erg\,s$^{-1}$, where another mechanism of X-ray production, probably magnetic in nature, dominates.

\section{BINARIES} 
We will here discuss only O+OB systems since no B+B binary had enough counts for a spectral study.

As can be seen from Table~\ref{ratio} and Figure~\ref{fig:lx}, the \lxlbol\ ratios of O+OB binaries appear quite similar to those of single O stars, with the exception of the bright system HD\,93403 \citep[a known colliding-wind binary, see][]{rau02}. This general lack of strong overluminosity was already found in \citet{osk05}, as well as for most binaries of NGC~6231 \citep{san06}, Carina~OB1 \citep{ant08}, and the 2XMM survey \citep{naz09}. It is confirmed here for the known massive binaries in the Carina Nebula. Thus there appear to be only a few exceptional systems where the collisions are X-ray-bright, the impact of wind-wind collision being small in most systems. Nevertheless, binaries and single objects are not 100\% equal. First, their average temperatures ($<kT>=\sum kT_i\times norm_i/\sum norm_i$) are slightly larger (for 1T fits: 0.47\,keV vs.\ 0.55\,keV for singles and binaries, respectively, for 2T fits: 0.71\,keV vs.\ 0.83\,keV for singles and binaries, respectively) but the differences are not significant (at the 1\% level), due to the large dispersion in average temperatures (up to 1\,keV!). Second, it must be noted that, whatever the energy band, the \loglxlbol\ values of binaries are systematically larger than those of single stars (as can be seen by looking at rows 4--5 vs.\ rows 9--10 of Table~\ref{ratio}). However, except for the medium energy band (which is hard enough to be potentially affected by some wind-wind emission), the differences are again not formally significant (at the 1\% level). 

Additional relations were also searched for in O+OB binaries, notably between \loglxlbol\ and the binary properties (total bolometric luminosity, eccentricity, period, semi-major axis; see Figure~\ref{fig:bin}). The most obvious trends are larger \loglxlbol\ ratios for longer periods, combined bolometric luminosity, or larger separation. However, these trends are clearly dominated by HD\,93403, the only known binary in the survey with period $P>10$\,days (the second binary of the QZ~Car system has a period of 20\,d, but the X-ray emission of QZ~Car cannot be related exclusively to that system; see \citealt{Parkin11}). Adding the binaries of NGC~6231 \citep{san06} totally blurs the picture. Indeed, for a large sample, \citet{naz09} found no strong relation (i.e., a very large scatter) between \loglxlbol\ and binary periods. Finally, while HD\,93403 stands out with its slightly harder emission and rather long period, the hardness ratios of the other O+OB binaries do not seem to correlate with their periods.

The small impact of wind-wind emission in the X-ray range can be qualitatively explained by two reasons. First, the winds collide at low speed in close binaries whereas the collision in wide systems is adiabatic, hence emitting few photons; there is thus a small region of parameter space where the conditions are just right for getting an X-ray-bright wind-wind collision -- this would be the case for HD\,93403. Second, it now seems difficult to get ram pressure balance in many systems where the wind momenta of the components are very different; in most cases, modeling shows that the stronger wind crashes onto its companion \citep{pit10}. To test these ideas quantitatively, comparisons of similar stellar pairs with, e.g., different periods are needed. However, although the Carina Nebula contains many objects, there are not enough cases of binaries to perform such a study. 

\section{WEAK-WIND OBJECTS}

The weak-wind problem in hot stars is twofold. On the one hand, some weak-wind stars have lower mass-loss rates than other stars of similar spectral type; on the other hand, other weak-wind stars have lower mass-loss rates than what atmosphere models predict. To explain the weakness of the winds, high X-ray luminosities have sometimes been invoked as they can modify the wind ionization, hence the efficiency of the wind acceleration \citep{mar05,mar09}. 

In our survey, six stars of the peculiar Vz type \citep{wal07} have been detected: HD\,93128, HD\,93129B, CPD\,--58$^{\circ}$2611, CPD\,--58$^{\circ}$2620, HD\,303311, and FO\,15. Adding HD\,93028 and possibly HD\,93146 from \citet{mar05}, this makes 8 stars with potential ``weak-wind'' problems. In our dataset, these 8 objects display normal total X-ray luminosities. A similar conclusion was found for HD\,46202, the only other weak-wind object which has been recently re-observed in X-rays: while old {\em Einstein} data seemed to imply a large \loglxlbol=--6, with a large impact on the ionization structure hence the wind propulsion \citep{mar05}, {\em Chandra} observations show the star to have \loglxlbol=--7 \citep{wan08}. Such ``canonical'' X-ray luminosities are probably insufficient to induce some wind-decreasing effect \citep[\loglxlbol=--3.5 might be necessary, see ][]{mar09}. It is however interesting to note that the \loglxlbol\ values in the soft and medium bands generally differ by 0.2--0.4\,dex (see Table~\ref{ratio}; only the supergiant O stars have a smaller difference but the sample size is just 2 objects), while this difference amounts to 0.6\,dex in these 8 stars. The X-ray emission of these potential ``weak-wind'' stars thus appears softer than the other stars of our sample although this is not a significant result (at the 1\% level, Figure~\ref{fig:weakw}). 

\section{THE FAINTEST DETECTED O AND B STARS} 

With at most 50 counts, the faintest O and B stars (8 O stars, 41 single B stars, and 2 B+B binaries) detected in this survey could not be analyzed spectroscopically on an individual basis. Stacking the data yields a composite spectrum for the single O stars detected with $<$50 counts and a separate composite spectrum for the single B stars detected with $<$50 counts.  Stacked spectra are always dominated by the strongest of the X-ray sources that were stacked, but we note that only about 10\% of these faint OB stars have substantially fewer counts than the others of the same category, limiting the consequences of stacking. These combined spectra were fitted by models similar to those described in Section~2 for individual stars, using an average interstellar column of 2.6$\times10^{21}$\,cm$^{-2}$ for the O stars and 2.2$\times10^{21}$\,cm$^{-2}$ for the B stars (which corresponds to the average interstellar column for these two stellar groups).  The parameters of the fits can be found at the bottom of Table~\ref{bestfit}.
These fits permit to find the conversion factor between photon fluxes and (unabsorbed) energy fluxes in the total band (the most reliable since it has the maximum number of counts): a unit photon flux corresponds to absorption-corrected energy fluxes of 4.05$\times10^{-9}$\,erg\,cm$^{-2}$\,s$^{-1}$ for O stars and 2.98$\times10^{-9}$\,erg\,cm$^{-2}$\,s$^{-1}$ for B stars. We then assume that the spectral parameters obtained from fitting the composite spectrum can be applied to each faint source individually. Table~\ref{faindetections} provides the X-ray source name and the stellar identification (Columns 1 and 2), the spectral type (Column 3), the binary status (Column 4), the number of net counts in the 0.5--8.0\,keV band together with their associated 68\% confidence interval (Column 5), the photon flux (Column 6), the median energy of the recorded counts (Column 7), the bolometric luminosity (Column 8), the absorption-corrected X-ray luminosity in the total energy band (Column 9), and the \lxlbol\ ratio in the same energy band (Column 10). Columns 5--7 are reproduced from the CCCP catalog \citep{Broos11}, columns 2--4 and 8 from the stellar catalog \citep{Povich11, Gagne11}. Figure~\ref{fig:faint} graphically illustrates the results. Error bars were estimated using the relative errors on the net counts; they are shown for demonstration purposes only (they are no perfect estimates of the actual errors since they do not take into account the modeling errors).  

For O stars, the 8 faint objects, all of late spectral type, clearly follow the \lxlbol\ relation traced by the brighter stars, confirming the above results. For B stars, however, the 41 faint stars do not behave as the other objects although all are of early-B spectral types (B0--B2). With similar bolometric luminosities but much lower X-ray luminosities, the faint objects simply populate the bottom part of the graphics. Although \loglxlbol\ still decreases with bolometric luminosity, it is obvious that the X-ray luminosity is thus not a constant for the majority of B stars. The B stars described in the above sections simply correspond to the tip of the iceberg, i.e., the X-ray-brightest cases. This range in X-ray luminosity implies a variety of emission processes in B stars; stellar winds and magnetic fields of various strengths, as well as different PMS companions in various states of flaring, could explain this variety of behaviors. 

Note that the same conversion factors were used to derive upper limits for the 71 undetected sources. Table~\ref{undetections} provides the position and the stellar identification of the source (Columns 1--3), the spectral type (Column 4), the binary status (Column 5), the upper limit on the number of net counts in the 0.5--8.0\,keV band (Column 6), the upper limit on the associated photon flux (Column 7), the bolometric luminosity (Column 8), the upper limit on the absorption-corrected X-ray luminosity (Column 9), and the upper limit on the \lxlbol\ ratio in the total energy band (Column 10). These upper limits correspond to the upper limit of the 68\% confidence interval, i.e., they are $\sim1\sigma$ upper limits. It is interesting to note, however, that the upper limit on the \lxlbol\ ratio is quite low for the two undetected O-type stars, being at least 1~dex below the average value. Without better knowledge of these two stars, this surprising result remains unexplained. The case of the undetected B stars is discussed in more detail in \citet{Evans11}.

\section{SUMMARY AND CONCLUSION}
The CCCP has detected the X-ray emission of 129 OB stars in the Carina Nebula (or about 65\% of the massive stars present in the field-of-view) . Among these, 78 sources display enough counts to be analyzed spectroscopically, at least roughly, and our {\em Chandra} observations therefore triple the number of massive stars studied in the Carina Nebula.

\lxlbol\ ratios (where X-ray luminosities are corrected for ISM absorption) were estimated for a large number of different cases. For single O-type stars, \loglxlbol\ is found to be --7.26 with a very small dispersion (0.2\,dex). There are only three clearly X-ray-overluminous objects: the binary HD\,93403 and the putative single stars HD\,93250 (a binary candidate) and Tr16\,MJ\,496 (a magnetic candidate?). No significant differences (at the 1\% level) in the ratio values are found when comparing single stars and binaries (although the ratios are systematically slightly larger for binaries, whatever the energy band considered), bright and faint stars, early and late stars, or main-sequence stars, giants, and supergiants. A trend of harder X-ray emission for the brighter stars is detected, but with a lot of scatter. Weak-wind stars appear to emit similar amounts of X-rays compared to ``normal'' O stars, but their X-ray emission may be slightly softer in character.  

For the few X-ray-bright B stars (i.e., those with $>$50 counts), the X-ray luminosity appears rather constant in the total and medium energy bands, and the spectra are harder than for O stars. When looking simultaneously at O and B stars, it seems that there is a soft transition from the latest-O to the X-ray bright, earliest-B stars. This may suggest that, for these few X-ray bright B-stars, an emission mechanism declines towards lower luminosities (the wind-shock model) while another one takes the lead (magnetic confinement?).

\acknowledgments

The authors thank J.\ McArthur for providing Oth order spectra of some of the piled-up sources, enabling comparison with the reconstructed data. The authors also thank the editor and the anonymous referee for their help. YN also thank her daughter Ana\"is C\'eleste for letting her finish this paper while she was only one week old. YN acknowledges support from the Fonds National de la Recherche Scientifique (Belgium), the PRODEX XMM and Integral contracts, and the `Action de Recherche Concert\'ee' (CFWB-Acad\'emie Wallonie Europe). LO acknowledges DLR grant 50\,OR\,0804 (LMO), JMP thanks the Royal Society for funding a University Research Fellowship, and DC acknowledges support from the {\em Chandra} grant G09-0019B. The Space Telescope Science Institute is operated by AURA, Inc., under NASA contract NAS5-26555. 
This work is supported by {\em {\em Chandra} X-ray Observatory} grant GO8-9131X (PI:  L.\ Townsley) and by the ACIS Instrument Team contract SV4-74018 (PI:  G.\ Garmire), issued by the {\em Chandra} X-ray Center, which is operated by the Smithsonian Astrophysical Observatory for and on behalf of NASA under contract NAS8-03060. 

{\it Facilities:} \facility{CXO (ACIS)}.

\clearpage

\begin{figure}
\includegraphics[width=9cm]{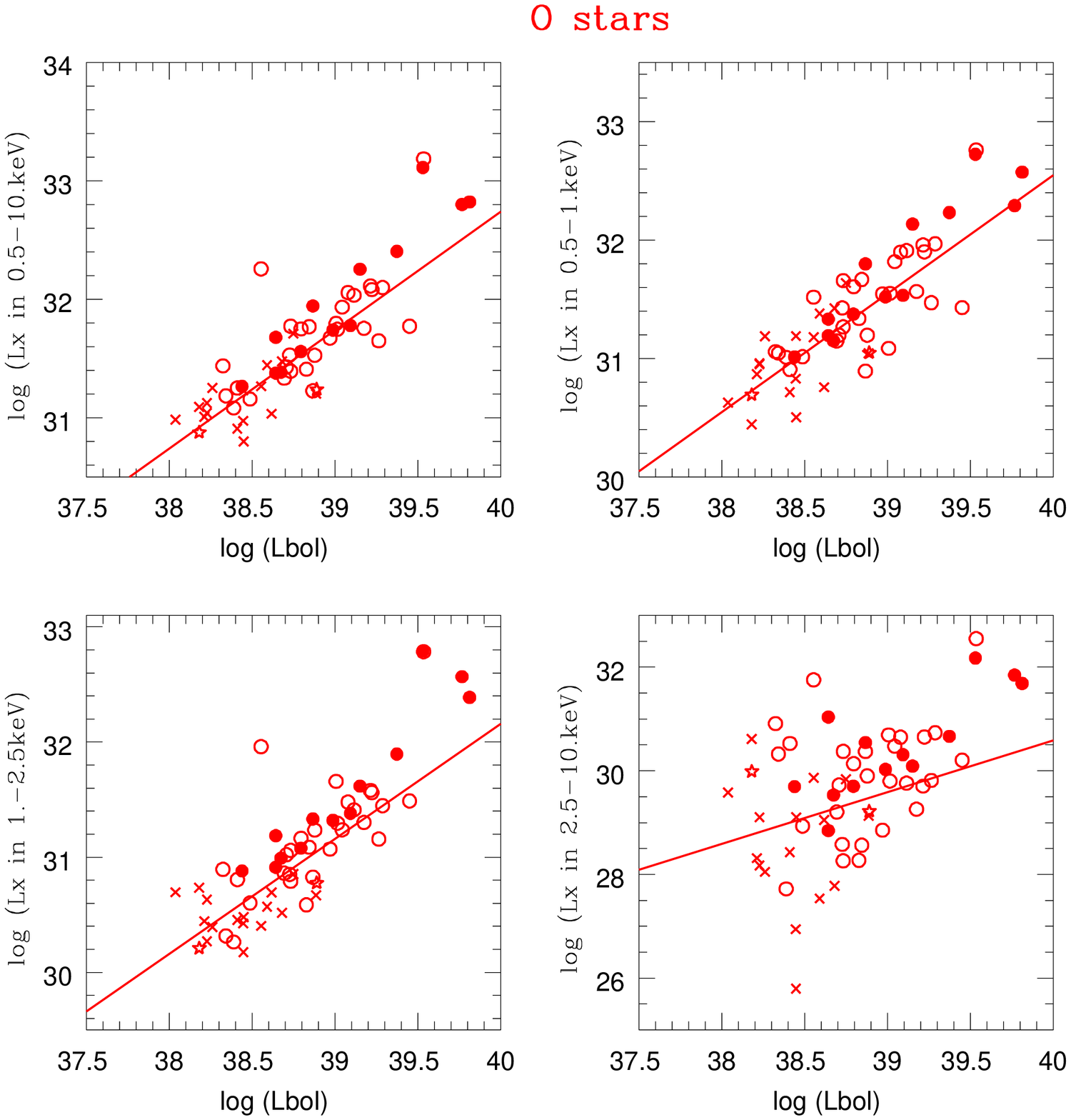}
\includegraphics[width=9cm]{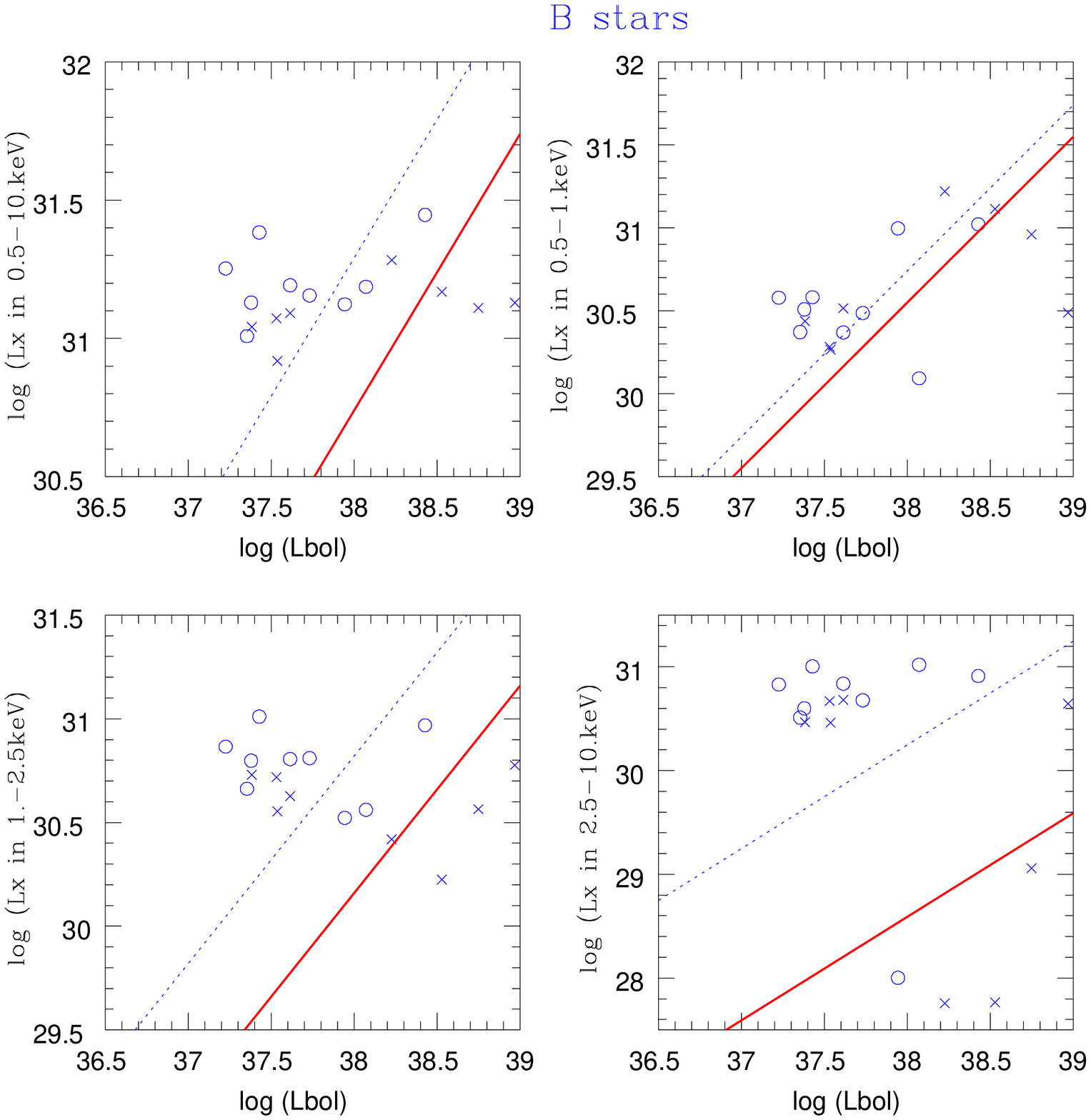}
\caption{X-ray luminosity in four energy bands as a function of bolometric luminosity, for O stars (upper 4 panels and lower 4 panels, in bold red), and B stars (middle 4 panels and lower 4 panels, in blue). Open and filled circles refer to singles and binaries, respectively, with $>$100 counts.  Crosses and stars are for singles and binaries recorded with only 50--100 counts. The lines give the \lxlbol\ relation from Table~\ref{ratio} derived for all single O (solid line) and B (dotted line) stars except HD\,93250 and Tr14\,MJ\,496. (A color version of this figure is available in the online journal.)  \label{fig:lx}}
\end{figure}
\setcounter{figure}{2}
\begin{figure}
\includegraphics[width=9cm]{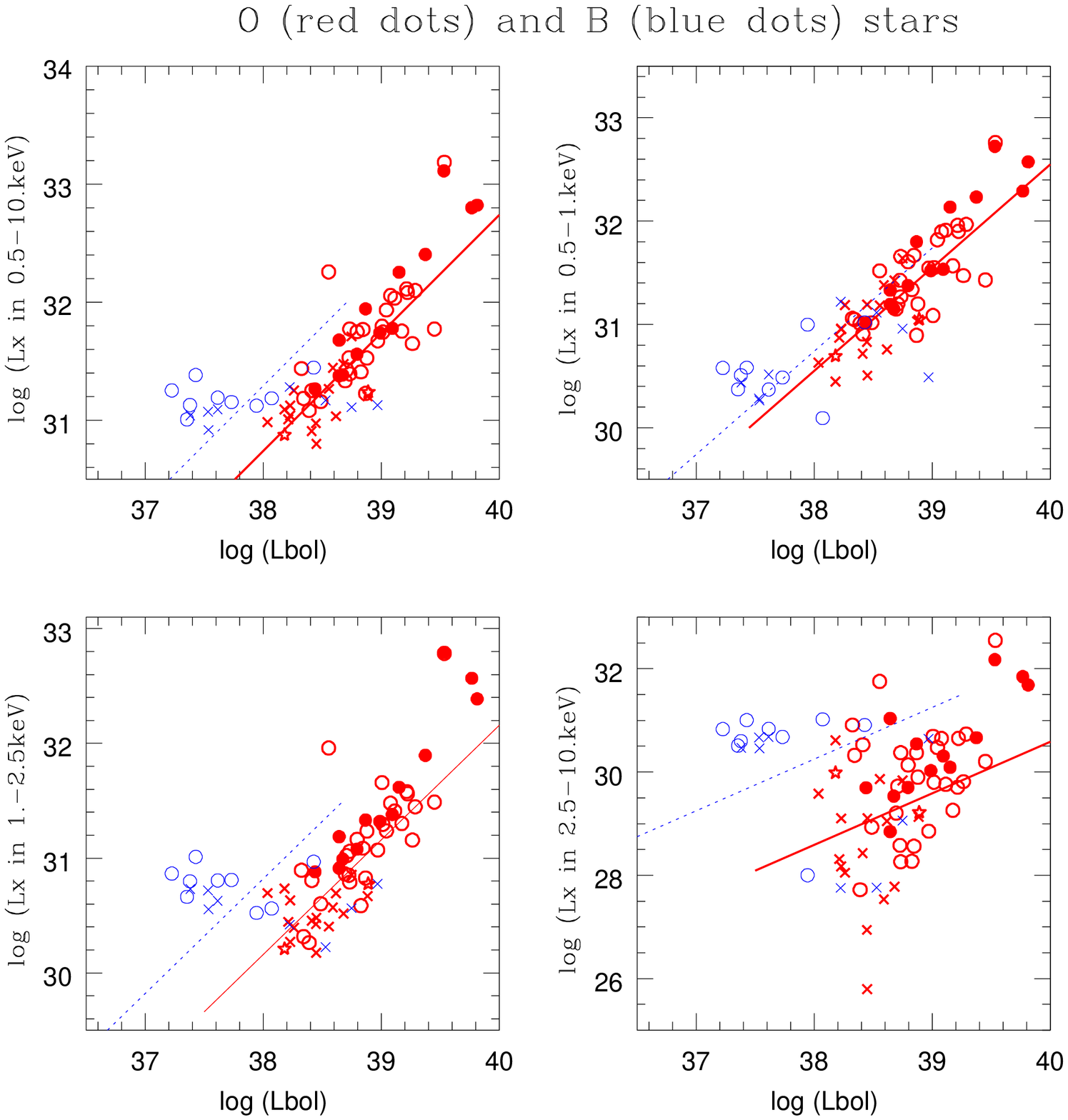}
\caption{Continued. }
\end{figure}

\begin{figure}
\includegraphics[width=9cm]{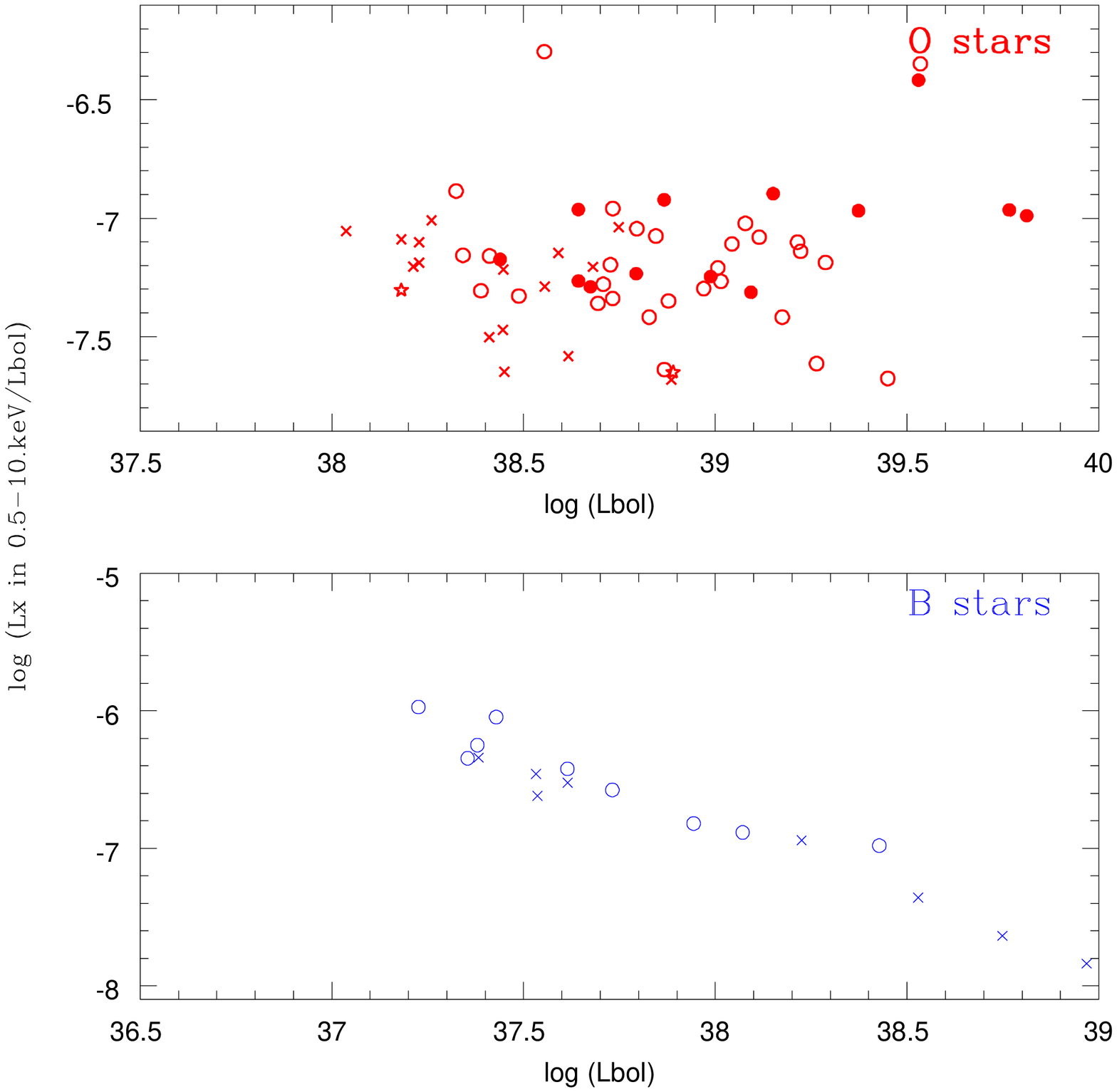}
\includegraphics[width=9cm]{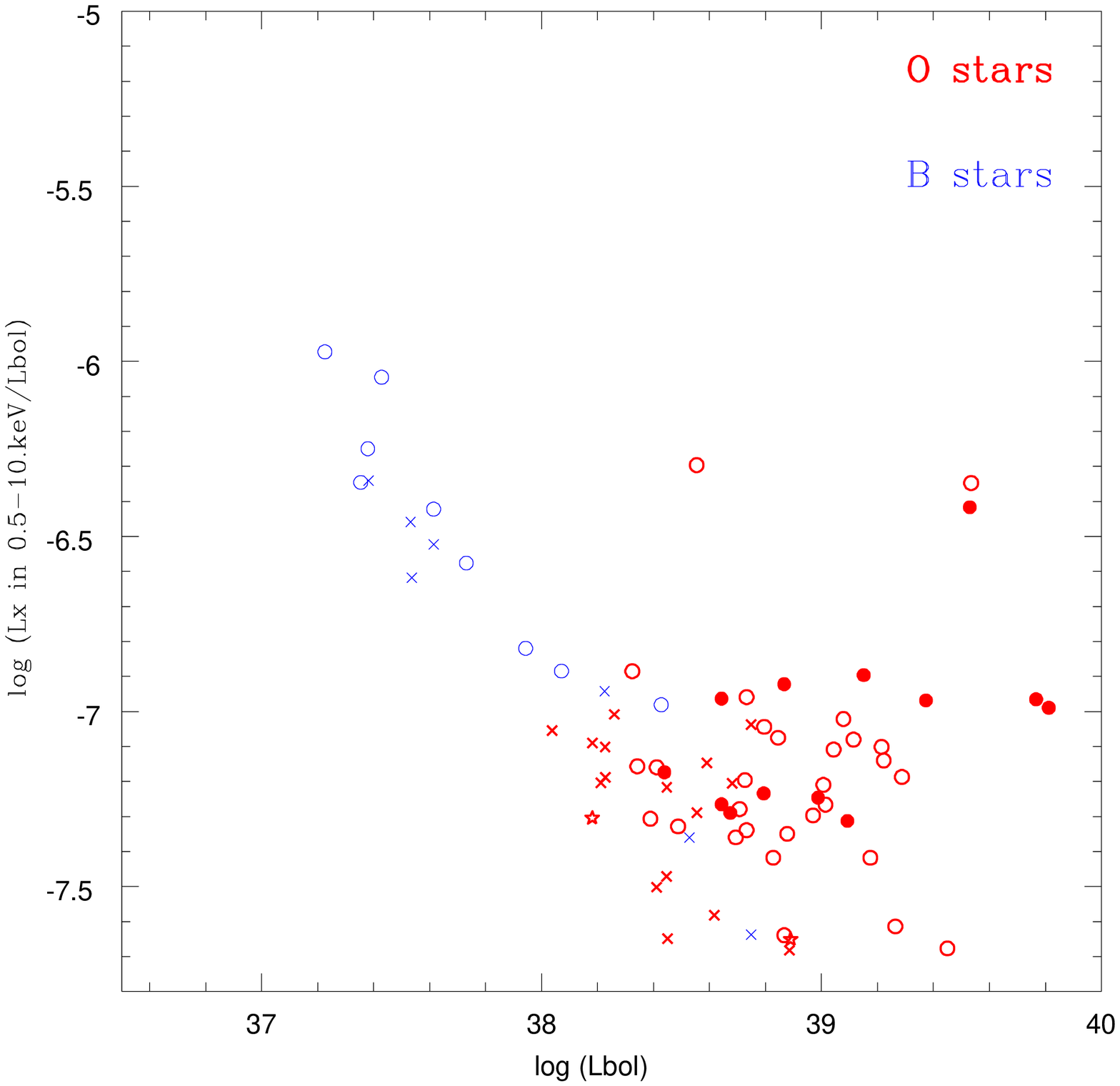}
\caption{\loglxlbol\ as a function of bolometric luminosity for the O (bold red) and B (blue) stars from our sample. Singles and binaries in two strata of detected counts are distinguished using the same symbols as in Figure~\ref{fig:lx}.  (A color version of this figure is available in the online journal.) \label{fig:ratio}}
\end{figure}

\begin{figure}
\includegraphics[width=9cm]{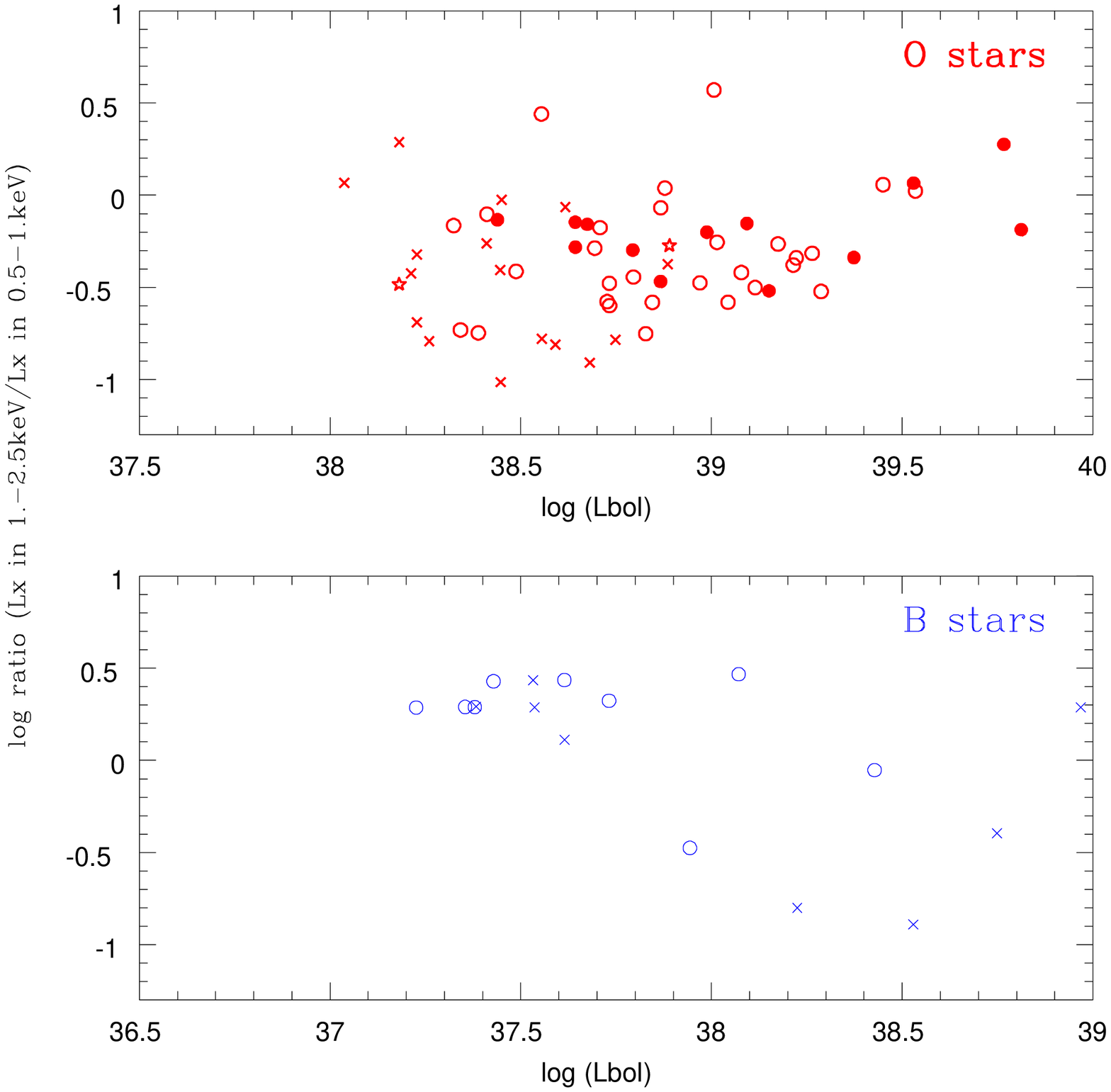}
\includegraphics[width=9cm]{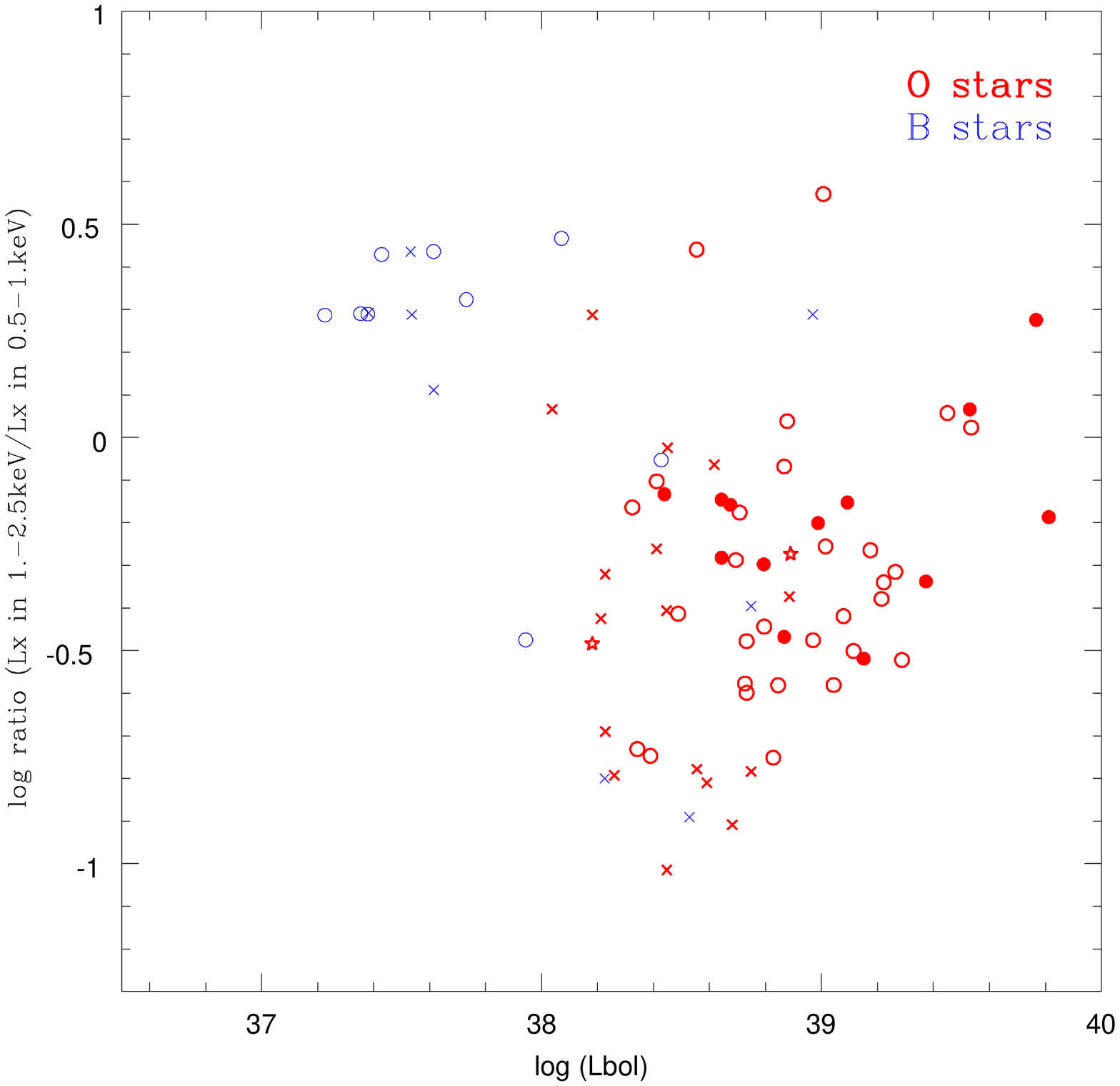}
\caption{Ratio of the medium and soft X-ray luminosities as a function of the bolometric luminosity for the O (bold red) and B (blue) stars from our sample. Singles and binaries in two strata of detected counts are distinguished using the same symbols as in Figure~\ref{fig:lx}. (A color version of this figure is available in the online journal.) \label{fig:hr}}
\end{figure}

\begin{figure}
\includegraphics[width=9cm]{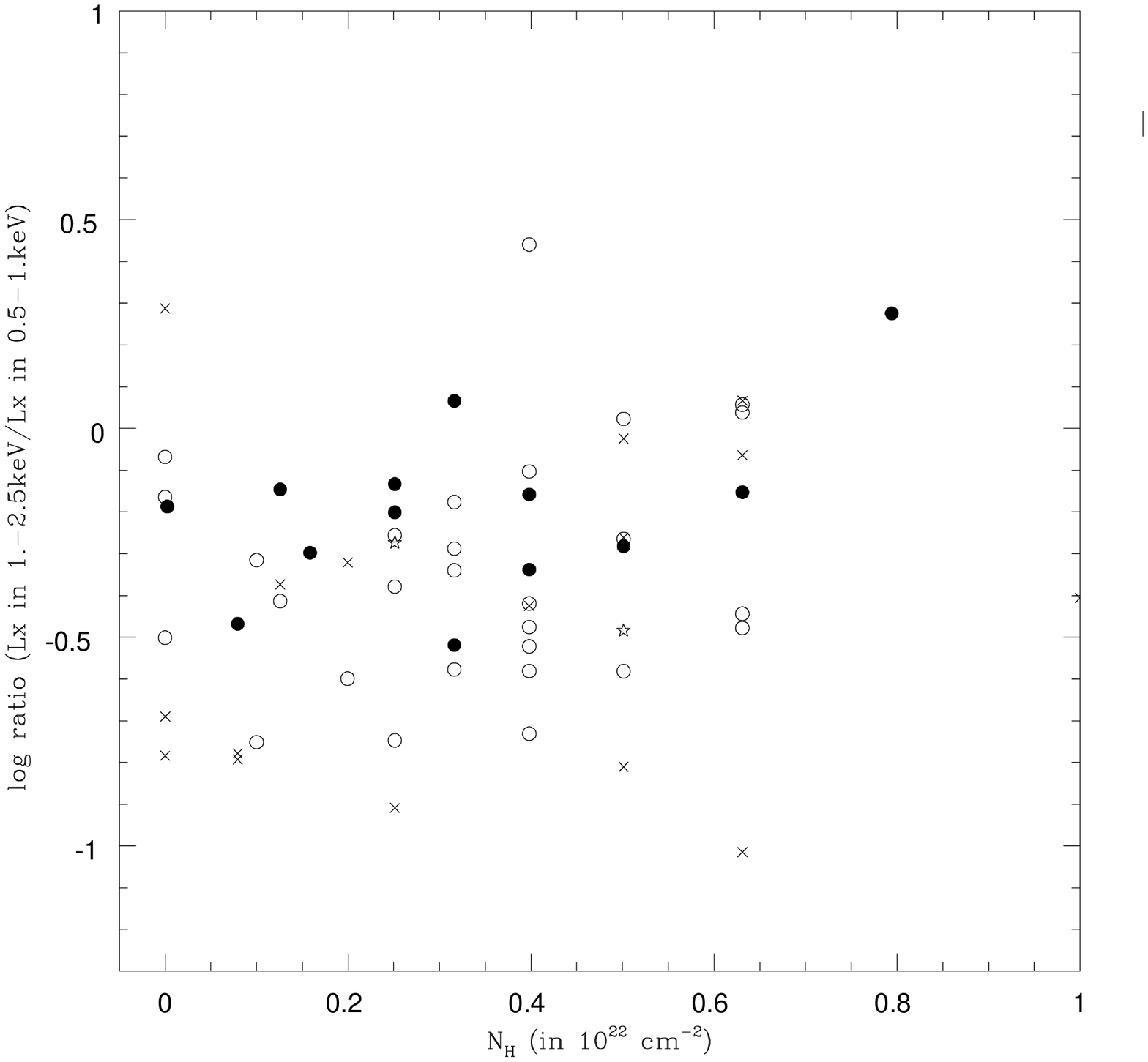}
\includegraphics[width=9cm]{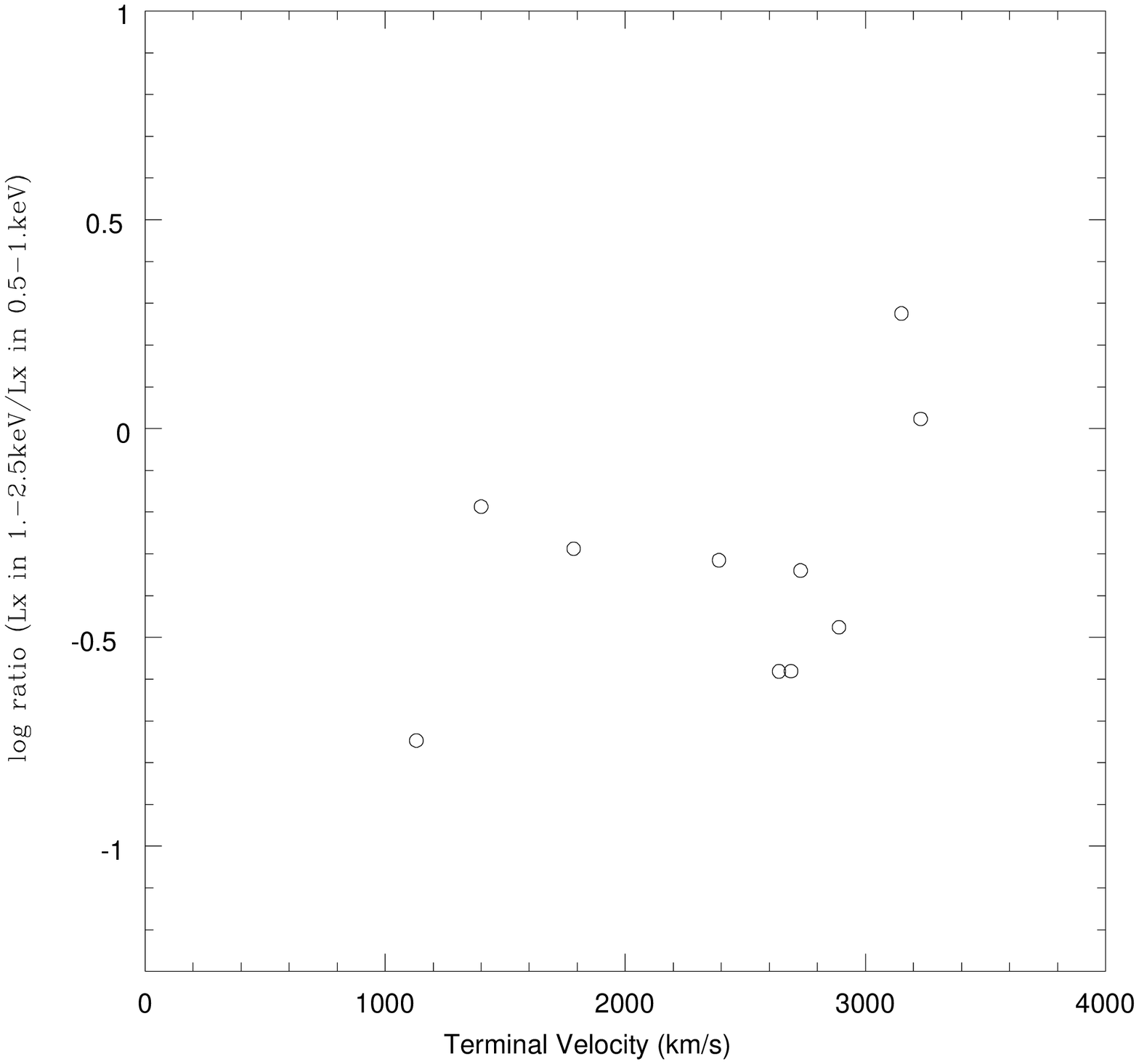}
\caption{Ratio of the medium and soft X-ray luminosities for O stars as a function of the additional absorbing column (top) and terminal wind velocity (when available, bottom). Symbols are as in Figure~\ref{fig:lx}. \label{fig:hr2}}
\end{figure}

\begin{figure}
\includegraphics[width=9cm]{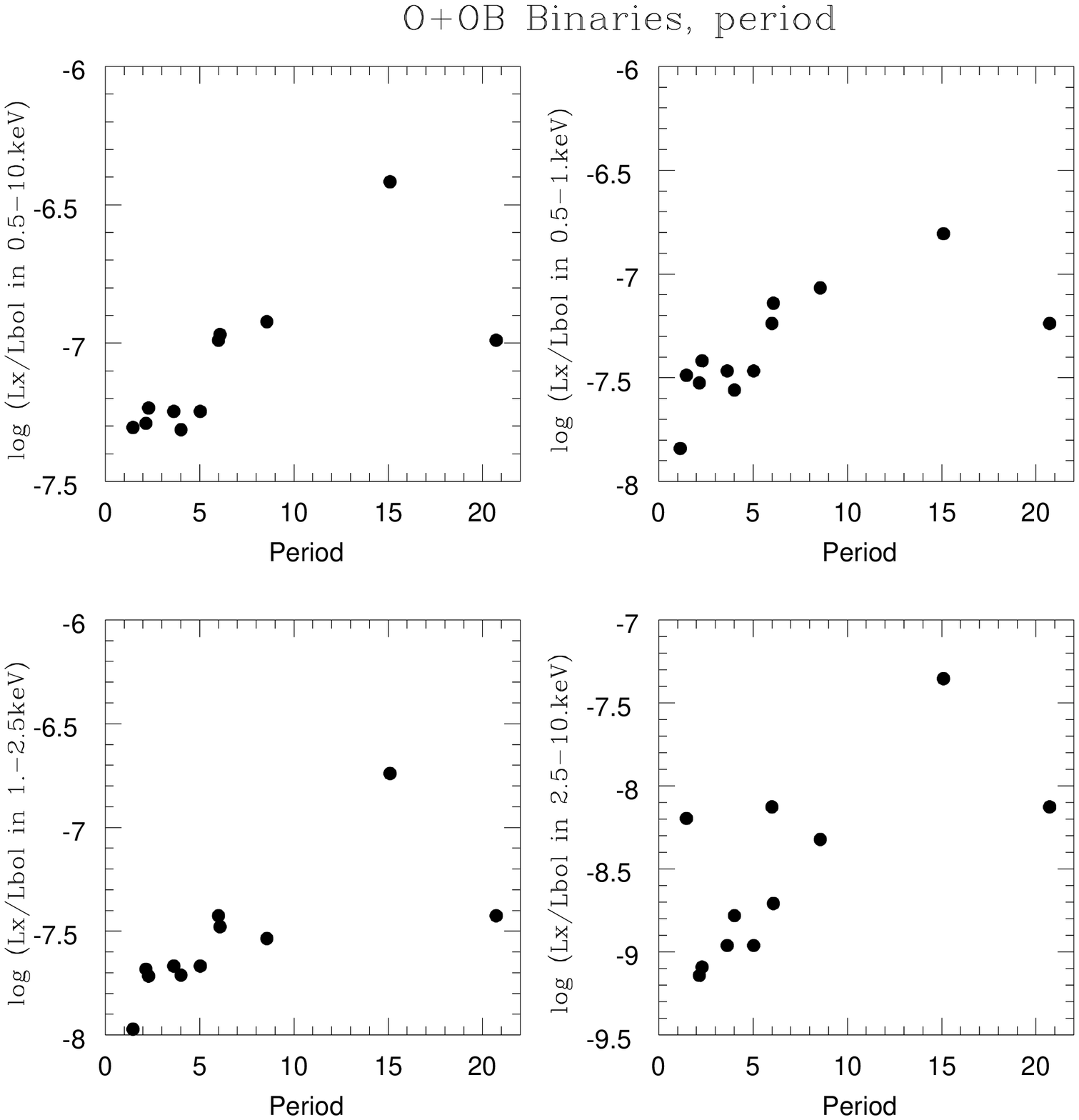}
\includegraphics[width=9cm]{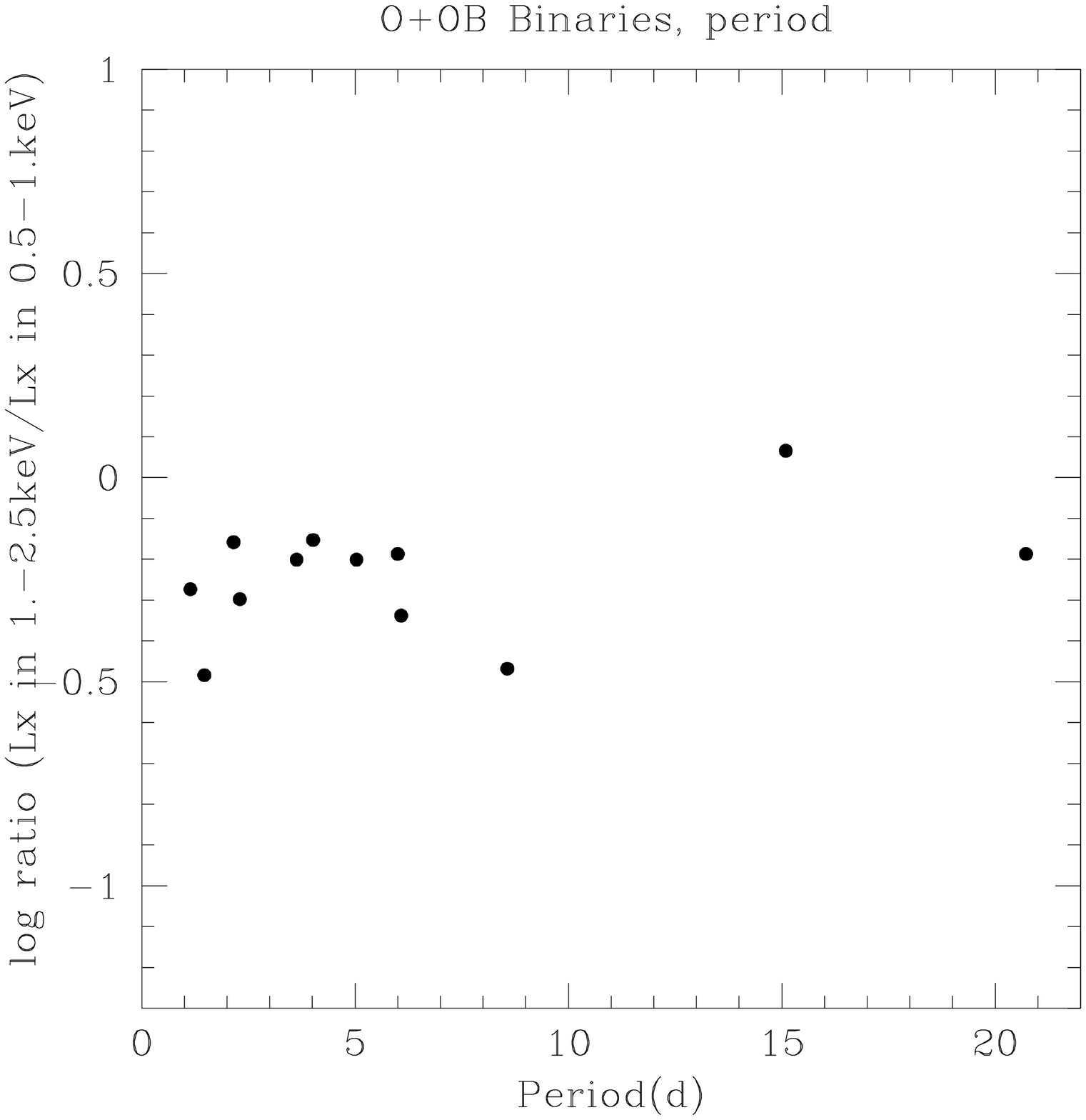}
\caption{\loglxlbol\ and hardness ratio as a function of the period (in days) for O+OB binaries. Note that a given object can appear twice if it is actually composed of two binary systems (e.g.\ QZ~Car). \label{fig:bin}}
\end{figure}

\begin{figure}
\includegraphics[width=9cm]{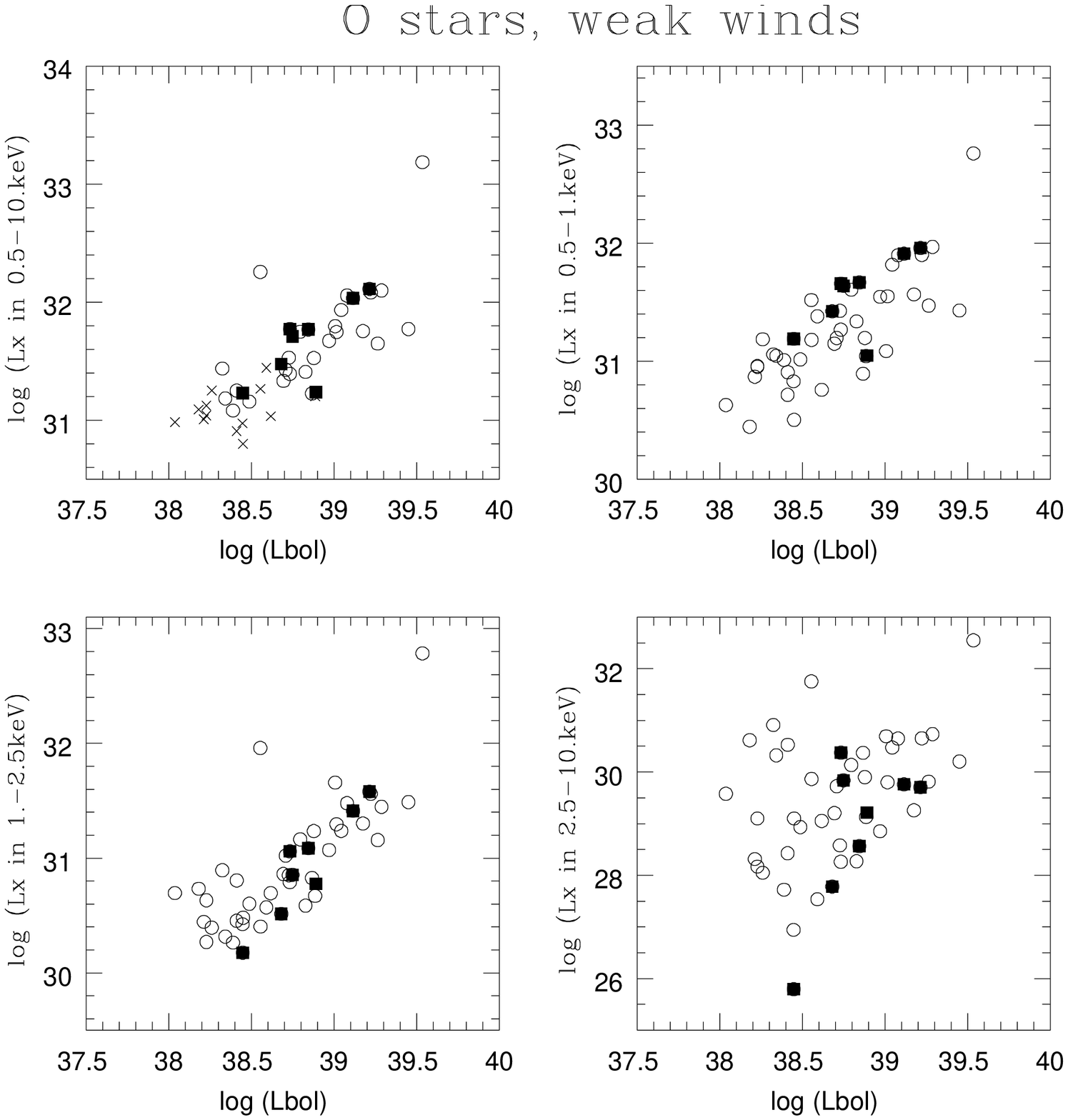}
\includegraphics[width=9cm]{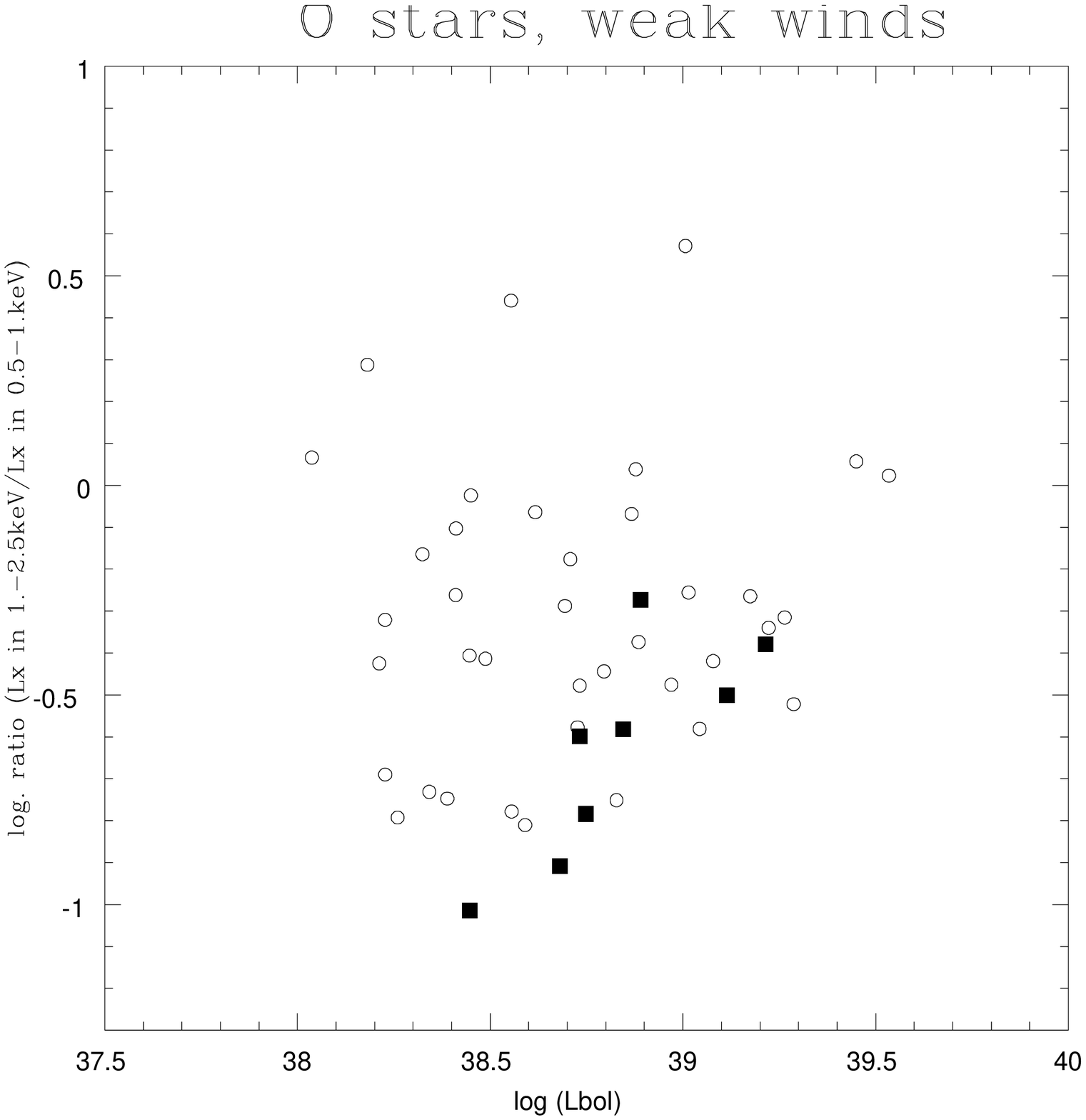}
\caption{Total X-ray luminosity (top) and ratio of the medium-to-soft fluxes (bottom) for weak-wind stars (filled squares) and ``normal'', single O stars (open circles). \label{fig:weakw}}
\end{figure}

\begin{figure}
\includegraphics[width=6.5cm]{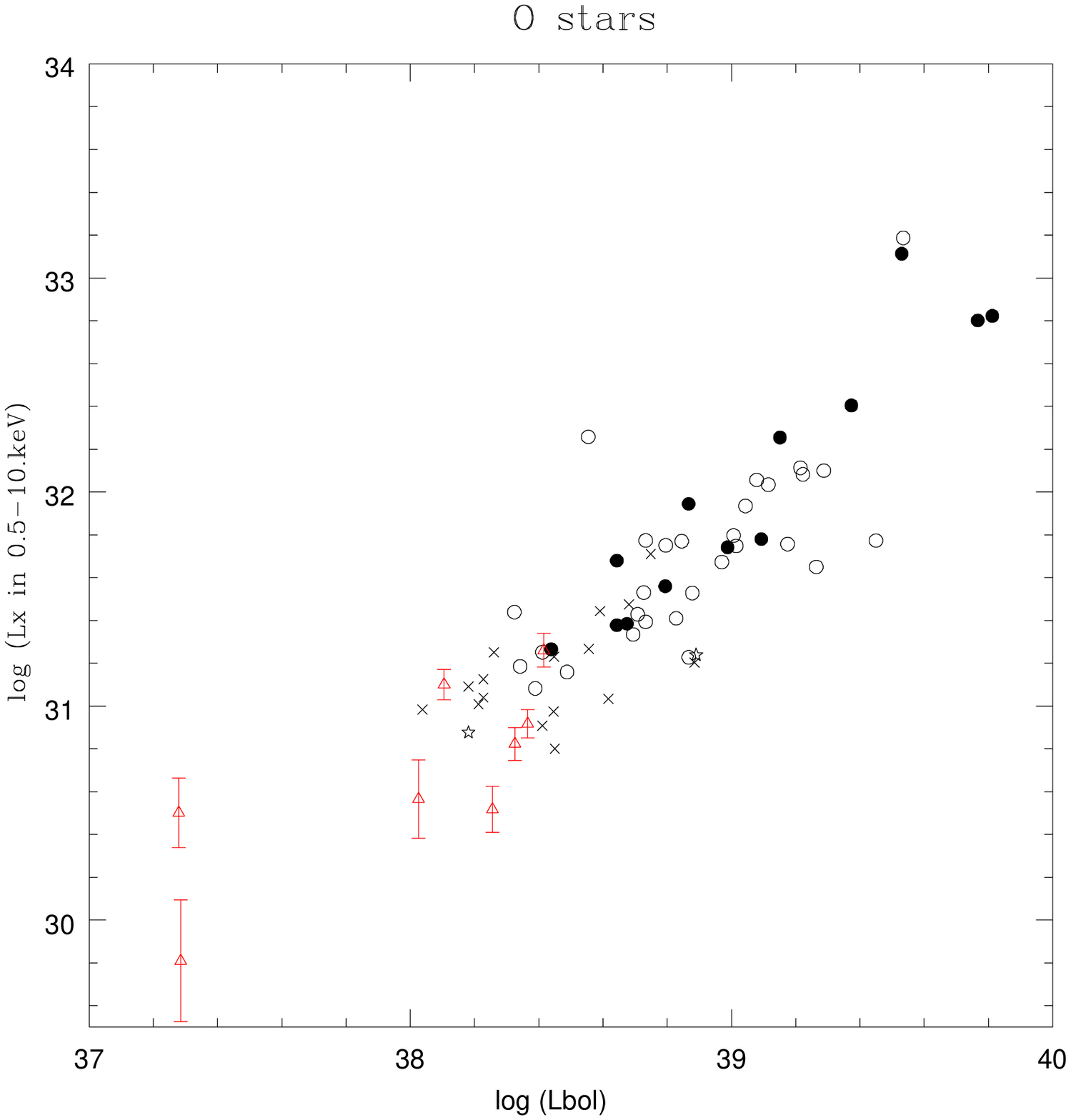}
\includegraphics[width=6.5cm]{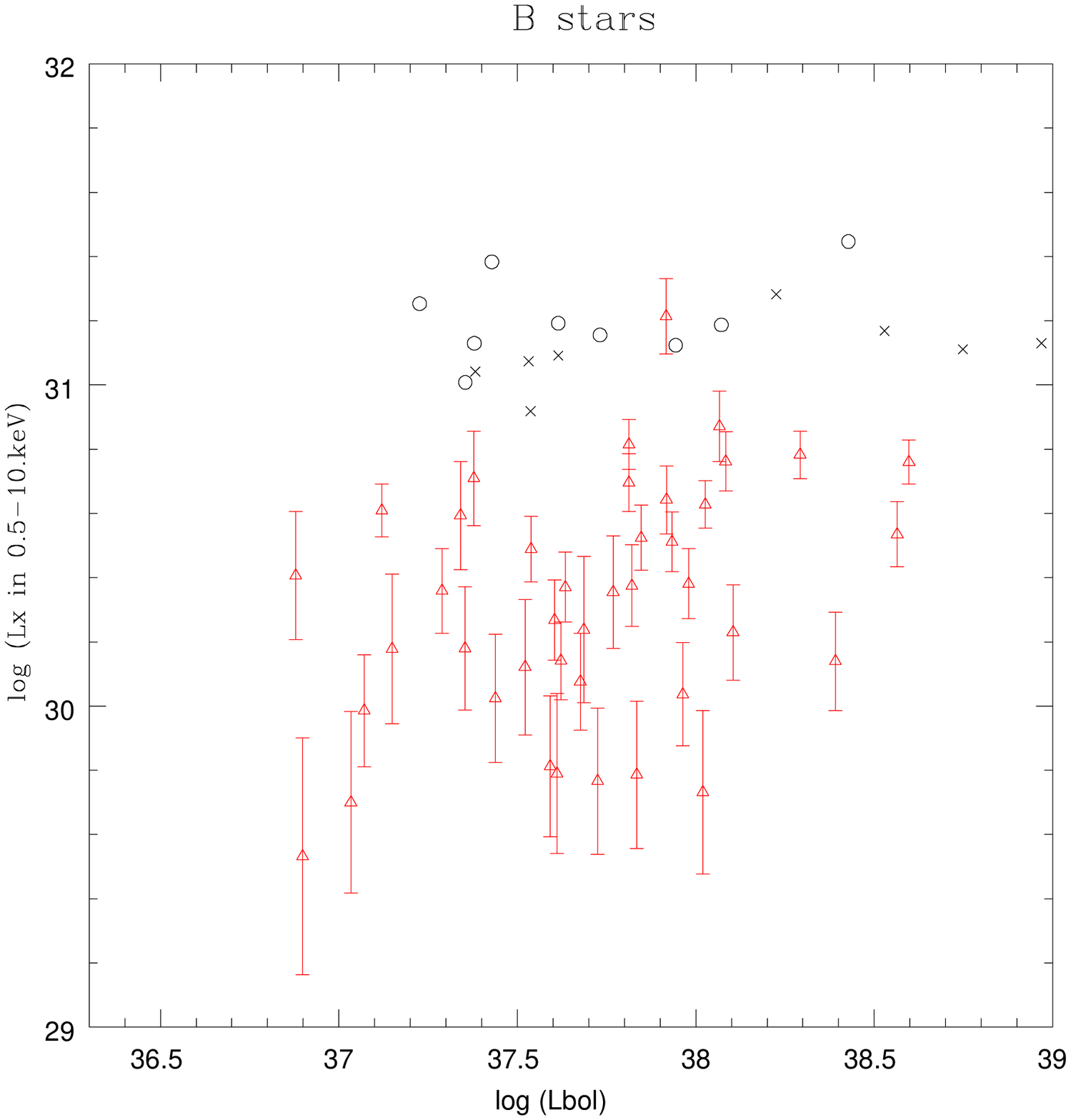}
\includegraphics[width=6.5cm]{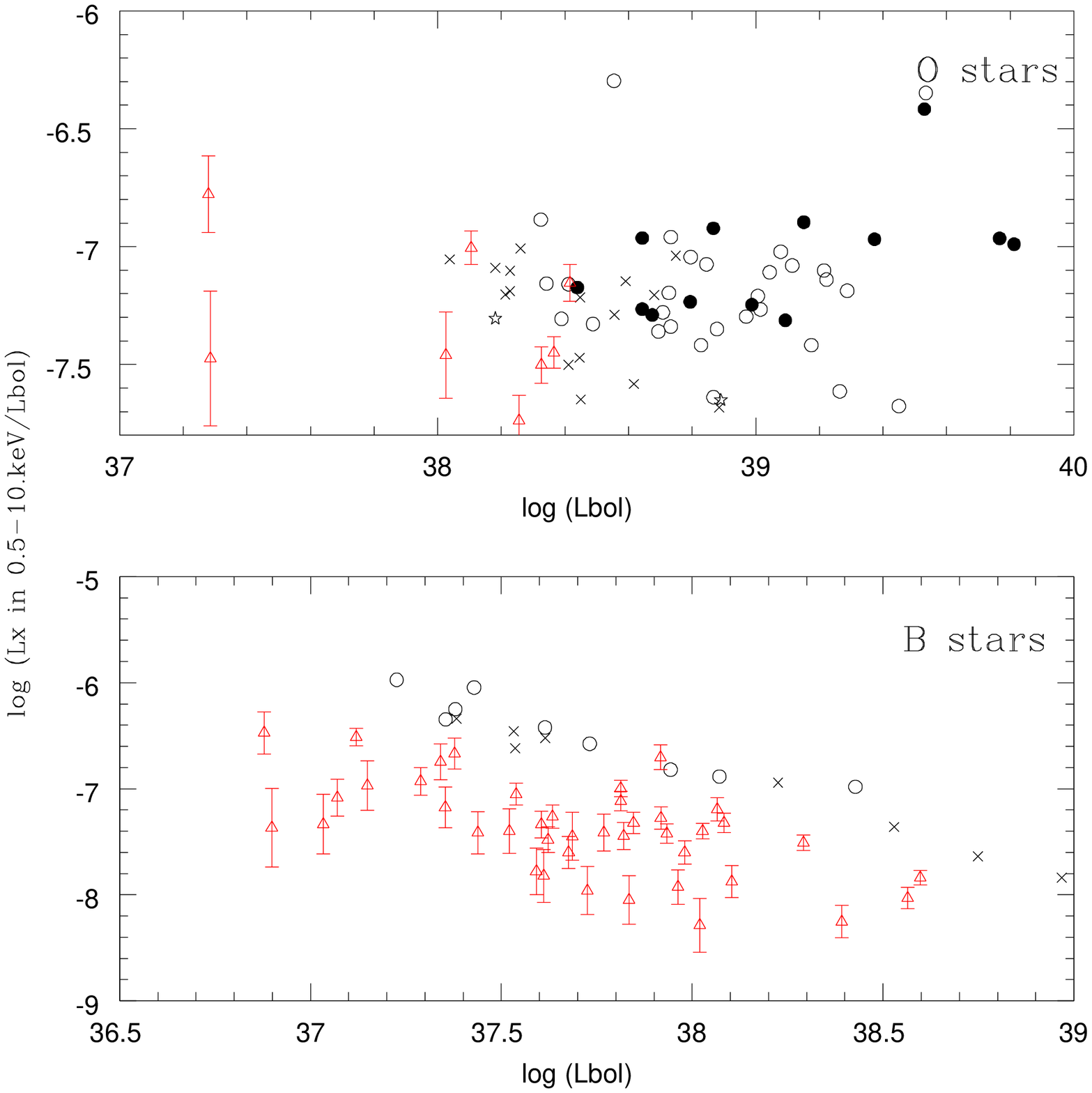}
\caption{X-ray luminosity and \loglxlbol\ as a function of the bolometric luminosity, for O stars and B stars. Singles and binaries in two strata of detected counts are distinguished using the same symbols as in Figure~\ref{fig:lx}. Faintest objects ($<$50 counts) in a third stratum of detected counts are shown by (red) triangles with error bars.  (A color version of this figure is available in the online journal.) \label{fig:faint}}
\end{figure}

\clearpage

\begin{table}
\begin{center}
\caption{Repartition of the detected hot stars as a function of the recorded counts. \label{repart}}
\begin{tabular}{l rrrr}
\tableline\tableline
 & O, single & O, binaries & B, single & B, binaries \\
\tableline
$>$100cts  & 29&13& 10& 0 \\
50--100cts & 16& 2&  8& 0 \\
$<$50cts   &  8& 0& 41& 2 \\
undetected &  2& 0& 67& 2 \\
\tableline
\end{tabular}
\end{center}
\end{table}

\clearpage
\renewcommand{\arraystretch}{1.5}
\setlength{\tabcolsep}{0.03in}
\begin{deluxetable}{rcccccccccccc}
\centering \rotate \tabletypesize{\tiny} \tablewidth{0pt}
\tablecaption{Best-fits results for the hot stars associated with X-ray sources with $>$50\,cts. Indices and exponents provides the lower and upper limits of the confidence interval (with 68\% confidence for counts and 90\% confidence for the spectral parameters and fluxes). \label{bestfit}}

\tablehead{
\colhead{CXOGNC J} & \colhead{Net Counts} & \colhead{Ph. Flux} & \colhead{Med. En.} & \colhead{$\chi^2_{\nu}$} & \colhead{dof} & \colhead{N$_H$(ISM)} & \colhead{$\log (N_H)$} & \colhead{kT$_1$} & \colhead{$\log EM_1$} & \colhead{kT$_2$} & \colhead{$\log EM_2$} & \colhead{Total Obs. Flux} \\
\colhead{} & \colhead{cts} & \colhead{ph cm$^{-2}$ s$^{-1}$} & \colhead{keV} & \colhead{} & \colhead{} & \colhead{10$^{22}$ cm$^{-2}$} & \colhead{} & \colhead{keV} & \colhead{} & \colhead{keV} & \colhead{} & \colhead{10$^{-14}$\,erg cm$^{-2}$ s$^{-1}$} \\
}

\startdata
104012.42-594810.1 & 416.8$_{396}^{437}$           &  3.39e-5        & 1.00 & 0.92 & 18 & 0.22 & $21.8\phd_{\cdots}^{+0.1}$ & ${\phn}0.18\phd_{-0.03}^{+0.13}$ & $56.2\phd_{-9.4}^{+0.3}$    & ${\phn}1.1\phd_{\cdots}^{+0.6}$ & $54.2\phd_{-0.3}^{+0.5}$    & 5.09$_{1.87}^{5.2}$ \\
104031.71-594643.9 & 131.1$_{120}^{143}$           &  9.55e-6        & 0.96 & 0.29 & 5  & 0.12 & $21.8\phd_{-0.4}^{+0.2}$   & ${\phn}0.24\phd_{-0.07}^{+0.15}$ & $55.2\phd_{\cdots}^{+0.9}$  & \nodata                     & \nodata                     & 1.53$_{0.}^{1.6}$   \\
104210.35-595800.9 & 64.2$_{56.3}^{72.7}$          &  5.40e-6        & 1.03 & 0.38 & 5  & 0.48 & $20.9\phd_{\cdots}^{+0.7}$ & ${\phn}0.56\phd_{-0.3}^{+0.2}$   & $54.0\phd_{-0.2}^{+0.7}$    & \nodata                     & \nodata                     & 0.74$_{0.18}^{0.85}$\\
104243.71-595416.6 & 106.6$_{95.2}^{118}$          &  6.40e-6        & 0.91 & 0.87 & 3  & 0.14 & $21.6\phd_{\cdots}^{+0.2}$ & ${\phn}0.20\phd_{\cdots}^{+0.1}$ & $55.0\phd_{-8.2}^{+0.4}$    & $10.\phd$                   & $53.3\phd_{-0.4}^{+0.4}$      & 1.64$_{0.18}^{1.45}$ \\
104311.17-594420.8 & 133.4$_{122}^{145}$           &  1.68e-5        & 0.97 & 0.39 & 5  & 0.37 & $21.5\phd_{\cdots}^{+0.3}$ & ${\phn}0.30\phd_{-0.1}^{+0.3}$   & $55.0\phd_{\cdots}^{+0.9}$  & \nodata                     & \nodata                     & 2.03$_{0.}^{2.19}$   \\
104315.33-601204.3 & 57.7$_{50.1}^{65.4}$          &  1.45e-5        & 0.90 & 1.09 & 5  & 0.13 & $21.8\phd_{-0.3}^{+0.2}$   & ${\phn}0.14\phd_{-0.03}^{+0.07}$ & $56.2\phd$                & \nodata                     & \nodata                     & 1.6$_{0.}^{1.54}$    \\
104317.92-600803.1 & 122.7$_{110}^{136}$           &  5.68e-6        & 0.91 & 0.96 & 5  & 0.17 & $21.4\phd_{\cdots}^{+0.4}$ & ${\phn}0.27\phd_{-0.1}^{+0.1}$   & $54.5\phd_{-0.6}^{+1.3}$   & \nodata                     & \nodata                      & 1.11$_{0.}^{1.2}$ \\
\enddata
\tablecomments{Source 104557.13-595643.1 corresponds to the star Cl* Trumpler 16 MJ 640 (or Hen 3-485 or SS73\_24), which is of type Be pec. Without precise knowledge of its photometry and other properties, its bolometric luminosity and its interstellar column could not be derived, which is why this source was not considered further. It is only shown here for completeness.\\
For more details on the stacked objects, see Sect. 7.
}
\end{deluxetable}

\clearpage

\begin{deluxetable}{rcccccccccccc}
\centering \rotate \tabletypesize{\tiny} \tablewidth{0pt}
\tablecaption{Summary of sources' properties.  \label{lxlbollist}}

\tablehead{
\colhead{CXOGNC J} & \colhead{Name} & \colhead{Sp. Type} & \colhead{Bin?} & \colhead{$\log(L_{BOL}/L_{\odot})$} & \multicolumn{4}{c}{$\log (L_X)$} &  \multicolumn{4}{c}{$\log (L_X/L_{BOL})$}\\
& & & & & \multicolumn{4}{c}{\hrulefill} & \multicolumn{4}{c}{\hrulefill}\\
\colhead{} & \colhead{} & \colhead{} & \colhead{} & \colhead{} & \colhead{total} & \colhead{soft} & \colhead{medium} & \colhead{hard} & \colhead{total} & \colhead{soft} & \colhead{medium} & \colhead{hard}\\
}

\startdata
104012.42-594810.1 & HD 92607               & O8 V                     &  & 5.21 & 31.75 & 31.61 & 31.16 & 30.14 & -7.04 & -7.19 & -7.63 &  -8.66 \\
104031.71-594643.9 & HD 92644               & B1.5 III                 &  & 4.36 & 31.12 & 31.00 & 30.52 & 28.00 & -6.82 & -6.95 & -7.42 &  -9.94 \\
104210.35-595800.9 & HD 305439A             & B0 Ia                    &  & 5.17 & 31.11 & 30.96 & 30.56 & 29.06 & -7.64 & -7.79 & -8.18 &  -9.69 \\
104243.71-595416.6 & HD 305438              & O8 V((f))                &  & 4.76 & 31.19 & 31.05 & 30.32 & 30.32 & -7.16 & -7.30 & -8.03 &  -8.02 \\
104311.17-594420.8 & HD 303316              & O6 V                     &  & 5.14 & 31.53 & 31.43 & 30.85 & 28.58 & -7.20 & -7.30 & -7.88 & -10.15  \\
104315.33-601204.3 & HD 93028               & O9 V                     &  & 4.86 & 31.23 & 31.19 & 30.18 & 25.80 & -7.22 & -7.26 & -8.27 & -12.65  \\
104317.92-600803.1 & HD 93027               & O9.5 V                   &  & 4.81 & 31.08 & 31.01 & 30.26 & 27.72 & -7.31 & -7.38 & -8.13 & -10.67  \\
104341.24-593548.1 & LS 1809                & O7 V                     &  & 5.03 & 31.03 & 30.76 & 30.69 & 29.05 & -7.58 & -7.86 & -7.92 &  -9.56 \\
\enddata
\tablecomments{Note that the X-ray luminosities have been corrected for the insterstellar absorption only (see text).}
\end{deluxetable}

\clearpage
\begin{table}
\begin{center}
\caption{\loglxlbol\ ratio. ("exc." means without the outliers - the single stars HD\,93250, Tr14\,MJ\,496, and/or the binary HD\,93403). Errors correspond to the 1-$\sigma$ dispersion of the data. \label{ratio}}
\begin{tabular}{l rrrrr}
\tableline\tableline
 & N & 0.5--10.\,keV & 0.5--1.\,keV & 1.--2.5\,keV & 2.5--10.\,keV \\
\tableline
{\bf O stars} \\
all, single                               & 45& $-7.22\pm0.28$ & $-7.42\pm0.28$ & $-7.79\pm0.33$ & $-9.30\pm1.17$\\
$>$100cts, single                         & 29& $-7.18\pm0.30$ & $-7.41\pm0.28$ & $-7.72\pm0.35$ & $-9.00\pm1.02$\\
50--100cts, single                        & 16& $-7.28\pm0.23$ & $-7.45\pm0.28$ & $-7.91\pm0.26$ & $-9.85\pm1.25$\\
\vspace*{-3mm}\\
all, single exc.                          & 43& $-7.26\pm0.21$ & $-7.45\pm0.26$ & $-7.84\pm0.23$ & $-9.41\pm1.07$\\
$>$100cts, single exc.                    & 27& $-7.24\pm0.20$ & $-7.45\pm0.25$ & $-7.80\pm0.21$ & $-9.15\pm0.87$\\
\end{tabular}
\end{center}
\end{table}

\clearpage
\begin{deluxetable}{rccccccccc}
\centering \rotate \tabletypesize{\tiny} \tablewidth{0pt}
\tablecaption{Properties of X-ray sources with $<$50\,cts. \label{faindetections}}

\tablehead{
\colhead{CXOGNC J} & \colhead{Name} & \colhead{Sp. Type} & \colhead{Bin?} & \colhead{$\log(L_{BOL}/L_{\odot})$} & \colhead{Net Counts} & \colhead{Ph. Flux} & \colhead{Med. En.} & \colhead{$\log (L_X)$} &  \colhead{$\log (L_X/L_{BOL})$}\\
\colhead{} & \colhead{} & \colhead{} & \colhead{} & \colhead{} & \colhead{cts} & \colhead{ph cm$^{-2}$ s$^{-1}$} & \colhead{} & \colhead{} \\

}

\startdata
104039.26-600536.1 & LS 1745                 & B2 III           &  & 3.76 & 16.0$_{10.1}^{22.5}$ & 2.08e-6 & 1.12  &  30.59  & -6.75  \\
104112.33-595825.0 & HD 92741                & B1.5 II:         &  & 4.98 & 23.8$_{18.6}^{29.6}$ & 1.82e-6 & 1.01  &  30.53  & -8.03  \\
104135.44-593945.6 & HD 303225               & B1.5 V           &  & 4.02 & 13.1$_{ 9.7}^{17.2}$ & 9.84e-7 & 1.07  &  30.27  & -7.34  \\
104213.35-590946.2 & HD 303297               & B1 V             &  & 4.23 & 38.3$_{31.7}^{45.4}$ & 3.47e-6 & 1.03  &  30.82  & -7.00  \\
104225.08-590924.7 & HD 303296               & B1 Ve            &  & 4.24 & 15.5$_{11.3}^{20.4}$ & 1.26e-6 & 1.00  &  30.38  & -7.45  \\	  
104236.31-595926.0 & CPD-59 2495             & B1.5 V           &  & 3.71 & 15.4$_{11.1}^{20.4}$ & 1.21e-6 & 1.42  &  30.36  & -6.93  \\  
104245.15-595219.5 & HD 305437               & B0.5 V           &  & 4.50 & 28.3$_{22.6}^{34.6}$ & 3.07e-6 & 0.85  &  30.76  & -7.32  \\	  
104303.96-595139.0 & HD 305515               & B1.5 Vsn:        &  & 3.77 &  8.6$_{ 5.1}^{12.7}$ & 8.02e-7 & 1.34  &  30.18  & -7.17  \\
\enddata
\tablecomments{Note that the X-ray luminosities have been corrected for the insterstellar absorption only (see text).}
\end{deluxetable}

\clearpage
\begin{deluxetable}{rccccccccc}
\centering \rotate \tabletypesize{\tiny} \tablewidth{0pt}
\tablecaption{Properties of the undetected OB stars. \label{undetections}}

\tablehead{
\colhead{RA} & \colhead{DEC} & \colhead{Name} & \colhead{Sp. Type} & \colhead{Bin?} & \colhead{$\log(L_{BOL}/L_{\odot})$} & \colhead{Up. Lim. on Net Cts} & \colhead{Up. Lim. on Ph. Flux} & \colhead{Up. Lim. on $\log (L_X)$} &  \colhead{Up. Lim. on $\log (L_X/L_{BOL})$}\\
\colhead{hhmmss} & \colhead{ddmmss} & \colhead{} & \colhead{} & \colhead{} & \colhead{} & \colhead{cts} & \colhead{ph cm$^{-2}$ s$^{-1}$} & \colhead{} & \colhead{} \\

}

\startdata
10 40 30.10 & -59 56 51.3 & HD 305443               & B2 III        &  & 3.71 & 5.7  & 5.46e-7 & 30.01   & -7.28  \\	
10 41 15.31 & -59 57 45.5 & LS 1760                 & B2 Ib         &  & 4.11 & 1.5  & 1.22e-7 & 29.36   & -8.33  \\	
10 41 20.28 & -60 06 36.4 & LS 1763                 & B2 V          &  & 3.54 & 1.4  & 1.12e-7 & 29.32   & -7.80  \\	
10 41 54.20 & -59 06 36.5 & HD 92852                & B1 V          &  & 4.16 & 7.5  & 7.16e-7 & 30.13   & -7.62  \\ 	
10 41 55.83 & -59 16 16.7 & HD 303202               & B3 III        &  & 3.80 & 2.8  & 2.36e-7 & 29.65   & -7.74  \\	
10 41 59.21 & -59 07 49.8 & HD 303189               & B2 V          &  & 3.70 & 1.2  & 2.55e-7 & 29.68   & -7.61  \\ 	
10 42 02.30 & -60 08 38.6 & HD 305452               & B2 III        &  & 4.11 & 2.5  & 2.04e-7 & 29.58   & -8.11  \\	
10 42 07.61 & -59 54 24.1 & HD 92877                & B2 III        &  & 4.22 &13.3  & 6.07e-7 & 30.06   & -7.74  \\
\enddata
\tablecomments{Note that the X-ray luminosities have been corrected for the insterstellar absorption only (see text).}
\end{deluxetable}


\begin{thebibliography}{}
\bibitem[Albacete Colombo et al.(2007)]{alb07} Albacete Colombo, J.~F., Flaccomio, E., Micela, G., Sciortino, S., \& Damiani, F.\ 2007, \aap, 464, 211 
\bibitem[Antokhin et al.(2008)]{ant08} Antokhin, I.~I., Rauw, G., Vreux, J.-M., van der Hucht, K.~A., \& Brown, J.~C.\ 2008, \aap, 477, 593 
\bibitem[Bergh\"ofer et al.(1997)]{ber97} Bergh\"ofer, T.W., Schmitt, J.H.M.M., Danner, R., \& Cassinelli, J.P. 1997, A\&A, 322, 167
\bibitem[Broos et al.(2010)]{Broos10} Broos, P.~S., Townsley, L.~K., Feigelson, E.~D., Getman, K.~V., Bauer, F.~E., \& Garmire, G.~P.\ 2010, \apj, 714, 1582 
\bibitem[Broos et al.(2011)]{Broos11} Broos, P.~S., et al.\ 2011, \apjs, submitted (CCCP Catalog Paper)
\bibitem[Carraro et al.(2004)]{car04} Carraro, G., Romaniello, M., Ventura, P., \& Patat, F.\ 2004, \aap, 418, 525 
\bibitem[Cassinelli \& Olson(1979)]{cas79} Cassinelli, J.~P., \& Olson, G.~L.\ 1979, \apj, 229, 304 
\bibitem[Chlebowski(1989)]{chl89} Chlebowski, T.\ 1989, \apj, 342, 1091 
\bibitem[Cohen et al.(1997)]{coh97} Cohen, D.~H., Cassinelli, J.~P., \& Macfarlane, J.~J.\ 1997, \apj, 487, 867 
\bibitem[Combi et al.(2011)]{combi} Combi, 2011, in Proceedings of the 39th LI\`ege Astrophysical Colloquium, eds. G. Rauw et al., BSRSL, vol 80, p
\bibitem[Cudworth et al.(1993)]{cud93} Cudworth, K.~M., Martin, S.~C., \& Degioia-Eastwood, K.\ 1993, \aj, 105, 1822 
\bibitem[DeGioia-Eastwood et al.(2001)]{deg01} DeGioia-Eastwood, K., Throop, H., Walker, G., \& Cudworth, K.~M.\ 2001, \apj, 549, 578 
\bibitem[Evans et al.(2004)]{eva04} Evans, N.~R., Schlegel, E.~M., Waldron, W.~L., Seward, F.~D., Krauss, M.~I., Nichols, J., \& Wolk, S.~J.\ 2004, \apj, 612, 1065 
\bibitem[Evans et al.(2011)]{Evans11} Evans, N.~R., et al.\ 2011, \apjs, submitted (CCCP Tr16 B~Stars Paper)
\bibitem[Gagn{\'e} et al.(2005)]{gag05} Gagn{\'e}, M., Oksala, M.~E., Cohen, D.~H., Tonnesen, S.~K., ud-Doula, A., Owocki, S.~P., Townsend, R.~H.~D., \& MacFarlane, J.~J.\ 2005, \apj, 628, 986 
\bibitem[Gagn{\'e} et al.(2011)]{Gagne11} Gagn{\'e}, M., et al.\ 2011, \apjs, submitted (CCCP Massive Star Signatures Paper)
\bibitem[Getman et al.(2005)]{get05} Getman, K. V., Feigelson, E. D., Grosso, N., McCaughrean, M. J., Micela, G., Broos, P., Garmire, G., \& Townsley, L. 2005, \apjs, 160, 353
\bibitem[Grillo et al.(1992)]{gri92} Grillo, F., Sciortino, S., Micela, G., Vaiana, G.~S., \& Harnden, F.~R., Jr.\ 1992, \apjs, 81, 795 
\bibitem[G{\"u}del et al.(2007)]{gue07} G{\"u}del, M., et al.\ 2007, \aap, 468, 353 
\bibitem[G{\"u}del \& Naz{\'e}(2009)]{gue09} G{\"u}del, M., \& Naz{\'e}, Y.\ 2009, \aapr, 17, 309 
\bibitem[Harnden et al.(1979)]{har79} Harnden, F.~R., Jr., et  al.\ 1979, \apjl, 234, L51 
\bibitem[Howarth et al.(1997)]{how97} Howarth, I.~D., Siebert, K.~W., Hussain, G.~A.~J., \& Prinja, R.~K.\ 1997, \mnras, 284, 265 
\bibitem[Lindgren(1968)]{lin68} Lindgren, B.W.\ 1968, Statistical theory - third edition, McMillan Pub. (New York)
\bibitem[Marcolino et al.(2009)]{mar09} Marcolino, W.~L.~F., Bouret, J.-C., Martins, F., Hillier, D.~J., Lanz, T., \& Escolano, C.\ 2009, \aap, 498, 837 
\bibitem[Martins et al.(2005)]{mar05b} Martins, F., Schaerer, D., \& Hillier, D.~J.\ 2005, \aap, 436, 1049 
\bibitem[Martins et al.(2005)]{mar05} Martins, F., Schaerer, D., Hillier, D.~J., Meynadier, F., Heydari-Malayeri, M., \& Walborn, N.~R.\ 2005, \aap, 441, 735 
\bibitem[Naz{\'e} et al.(2008)]{naz08} Naz{\'e}, Y., Rauw, G., \& Manfroid, J.\ 2008, \aap, 483, 171 
\bibitem[Naz{\'e}(2009)]{naz09} Naz{\'e}, Y.\ 2009, \aap, 506, 1055 
\bibitem[Naz{\'e}(2011)]{naz10} Naz{\'e}, Y.\ 2011, in Proceedings of the 39th LI\`ege Astrophysical Colloquium, eds. G. Rauw et al., BSRSL, vol 80, p109
\bibitem[Nelan et al.(2004)]{nel04} Nelan, E.~P., Walborn, N.~R., Wallace, D.~J., Moffat, A.~F.~J., Makidon, R.~B., Gies, D.~R., \& Panagia, N.\ 2004, \aj, 128, 323
\bibitem[Oskinova(2005)]{osk05} Oskinova, L.~M.\ 2005, \mnras, 361, 679 
\bibitem[Owocki \& Cohen(1999)]{owo99} Owocki, S.~P., \& Cohen, D.~H.\ 1999, \apj, 520, 833 
\bibitem[Pallavicini et al.(1981)]{pal81} Pallavicini, R., Golub, L., Rosner, R., Vaiana, G.S., Ayres, T., \& Linsky, J.L.\ 1981, \apj, 248, 279
\bibitem[Parkin et al.(2011)]{Parkin11} Parkin, E.~R., et al.\ 2011,Ê\apjs, submitted (CCCP QZ~Car Paper)
\bibitem[Pittard \& Parkin(2010)]{pit10} Pittard, J.M., \& Parkin, E.R.\ 2010, \mnras, 463, 1657
\bibitem[Povich et al.(2011)]{Povich11} Povich, M.~S., et al.\ 2011,Ê\apjs, submitted (CCCP Massive Star Candidates Paper)
\bibitem[Rauw et al.(2002)]{rau02} Rauw, G., Vreux, J.-M., Stevens, I.~R., Gosset, E., Sana, H., Jamar, C., \& Mason, K.~O.\ 2002, \aap, 388, 552 
\bibitem[Rauw et al.(2009)]{rau09} Rauw, G., Naz{\'e}, Y., Fern{\'a}ndez Laj{\'u}s, E., Lanotte, A.~A., Solivella, G.~R., Sana, H., \& Gosset, E.\ 2009, \mnras, 398, 1582 
\bibitem[Sana et al.(2006)]{san06} Sana, H., Rauw, G., Naz{\'e}, Y., Gosset, E., \& Vreux, J.-M.\ 2006, \mnras, 372, 661 
\bibitem[Sanz-Forcada et al.(2004)]{san04} Sanz-Forcada, J., Franciosini, E., \& Pallavicini, R.\ 2004, \aap, 421, 715 
\bibitem[Seward et al.(1979)]{sew79} Seward, F.~D., Forman, W.~R., Giacconi, R., Griffiths, R.~E., Harnden, F.~R., Jr., Jones, C., \& Pye, J.~P.\ 1979, \apjl, 234, L55 
\bibitem[Skiff(2009)]{ski09} Skiff\ 2009, ``Catalogue of Stellar Spectral Classifications,'' VizieR Catalog B/mk/mktypes
\bibitem[Smith et al.(2001)]{smi01} Smith, R.K., Brickhouse, N.S., Lieadhl, D.A., \& Raymond J.C.\ 2001, \apjl, 556, L91
\bibitem[Stelzer et al.(2005)]{ste05} Stelzer, B., Flaccomio, E., Montmerle, T., Micela, G., Sciortino, S., Favata, F., Preibisch, T., \& Feigelson, E.~D.\ 2005, \apjs, 160, 557 
\bibitem[Townsley et al.(2003)]{tow03} Townsley, L.~K., Feigelson, E.~D., Montmerle, T., Broos, P.~S., Chu, Y.-H., \& Garmire, G.~P.\ 2003, \apj, 593, 874 
\bibitem[Townsley et al.(2011)]{Townsley11} Townsley, L.~K., et al.\ 2011, \apjs, submitted (CCCP Intro Paper)
\bibitem[ud-Doula \& Owocki(2002)]{udd02} ud-Doula, A., \& Owocki, S.~P.\ 2002, \apj, 576, 413 
\bibitem[ud-Doula et al.(2006)]{udd06} ud-Doula, A., Townsend, R.~H.~D., \& Owocki, S.~P.\ 2006, \apjl, 640, L191 
\bibitem[Vuong et al.(2003)]{vuo03} Vuong, M. H., Montmerle, T., Grosso, N., Feigelson, E. D., Verstraete, L., \& Ozawa, H. 2003, A\&A, 408, 581
\bibitem[Vink et al.(2000)]{vin00} Vink, J.~S., de Koter, A., \& Lamers, H.~J.~G.~L.~M.\ 2000, \aap, 362, 295 
\bibitem[Walborn(1995)]{wal95} Walborn, N.~R.\ 1995, Revista Mexicana de Astronomia y Astrofisica Conference Series, 2, 51 
\bibitem[Walborn(2009)]{wal07} Walborn, N.~R.\ 2009, in STScI Symposium Series, Massive Stars from Pop. III and GRBS to the Milky Way (M. Livio and E. Villaver eds.), Vol. 20, 167 (arXiv:astro-ph/0701573) 
\bibitem[Walborn et al.(2009)]{wal09} Walborn, N.~R., Nichols, J.~S., \& Waldron, W.~L.\ 2009, \apj, 703, 633 
\bibitem[Wang et al.(2008)]{wan08} Wang, J., Townsley, L.~K., Feigelson, E.~D., Broos, P.~S., Getman, K.~V., Rom{\'a}n-Z{\'u}{\~n}iga, C.~G., \& Lada, E.\ 2008, \apj, 675, 464 
\bibitem[Wilms et al.(2000)]{wil00} Wilms, J., Allen, A., \& Mc Cray, R.\ 2000, \apj, 542, 914
\end{thebibliography}
\end{document}